\documentclass[reprint,aps,prb,floatfix,twocolumn,superscriptaddress,showpacs,amsmath,amssymb,showpacs,]{revtex4-1}
\usepackage{graphicx}%
\usepackage{bm}%
\usepackage{amsmath, nccmath}
\usepackage[USenglish]{babel}
\usepackage[utf8]{inputenc}

\begin{document}

\author{Viktor Christiansson}
\affiliation{Department of Physics, University of Fribourg, 1700 Fribourg, Switzerland}
\author{Francesco Petocchi} 
\affiliation{Department of Physics, University of Fribourg, 1700 Fribourg, Switzerland}
\author{Philipp Werner }
\affiliation{Department of Physics, University of Fribourg, 1700 Fribourg, Switzerland}

\title{$GW$+EDMFT investigation of Pr$_{1-x}$Sr$_x$NiO$_2$ under pressure}

\begin{abstract}

Motivated by the recent experimental observation of a large pressure effect on $T_c$ in Pr$_{1-x}$Sr$_x$NiO$_2$, we study the electronic properties of this compound as a function of pressure for $x=0$ and $0.2$ doping using self-consistent $GW$+EDMFT. Our numerical results demonstrate a non-trivial interplay between chemical doping and physical pressure, and small but systematic changes in the orbital occupations, local level energies, and interaction parameters with increasing pressure. The proper treatment of correlation effects, beyond density function theory, is shown to play an important role in revealing these trends. While the pressure dependent changes in the electronic structure of the undoped compound suggest a more single-band-like behavior in the high-pressure regime, a qualitatively different behavior is found in the doped system. We also point out that the fluctuations in the orbital occupations and spin states are not consistent with a single-band picture, and that at least a two-band model is necessary to reproduce the full result. This multi-orbital nature manifests itself most clearly in the doped compound.
\end{abstract}

\maketitle

\section{\label{sec:Introduction}Introduction}

Since the recent discovery of superconductivity in the infinite-layered phase of Sr-doped NdNiO$_2$,\cite{Li2019} an intense research effort, both on the experimental and theoretical side, has been devoted to understanding the electronic properties of this class of materials. In particular, the similarities and differences to the cuprate superconductors have received significant attention. In the mean time, the family of infinite-layered nickelates exhibiting superconductivity has grown to include Sr-doped PrNiO$_2$,\cite{Osada2020a,Osada2020b} and both Sr-\cite{Osada2021} and Ca-doped \cite{Zeng2022} LaNiO$_2$. Furthermore, superconductivity has recently been discovered in finite-layered nickelates ($n=5$ layers), in the absence of chemical doping,\cite{Pan2021} which provides an interesting additional avenue for exploring the pairing mechanism in this class of materials.

A widely debated but still not fully settled question concerns the single- versus multi-orbital nature of these systems. 
Some groups\cite{Nomura2019,Kitatani2020,Karp2020a,Karp2020b,Higashi2021} argue that the Ni $3d_{x^2-y^2}$ orbital is the main player and relevant for the observed superconductivity (as in the cuprates), while other groups\cite{Lechermann2020a,Werner2020,Lechermann2020b,Petocchi2020b,Kang2020,Wang2020} claim that the inclusion of additional orbitals is necessary to accurately describe the low-energy physics.
Other open questions range from the importance of the self-doping caused by the rare-earth atom layer separating the NiO$_2$ planes\cite{Lee2004,Gu2020} to the importance of the in-plane oxygen orbitals.\cite{Karp2020a} A recent review of the present understanding can be found, e.g., in Ref.~\onlinecite{Nomura2022}.

Recently, Wang {\it et al.}\cite{Wang2022} reported that upon applying pressure to doped Pr$_{1-x}$Sr$_x$NiO$_2$ ($x=0.18)$, the superconducting critical temperature $T_c$ is enhanced from 18~K at ambient conditions to 31~K at 12.1 GPa, with no sign of saturation. This opens up an interesting prospect for experimentally reaching even higher values of $T_c$, similar to the cuprates where the highest $T_c$ values are also reached under pressure.\cite{Gao1994,Monteverde2005} 
On the theoretical side the systematic trend with pressure provides a potentially fruitful venue 
to gain insights into the single- versus multi-orbital question, and ultimately a deeper understanding of the underlying mechanism of superconductivity in the nickelates. 

Some authors have considered an equivalent chemical pressure effect by altering the chemical composition leading to a change also in the $c$-lattice constant. In particular, a change in the rare-earth\cite{Been2021,Bernardini2022} in $R$NiO$_2$ was shown to result in large changes (Ref.~\onlinecite{Bernardini2022} finds that the change from La to Y corresponds to a pressure of $\sim19$ GPa), while additionally intercalating the structure with topotactic H in $R$NiO$_2$H results in small effects on the $c$-lattice parameter compared to the stoichiometric compound.\cite{Si2020,Malyi2022}
However, to our knowledge, only Been {\it et al.},\cite{Been2021} in their LDA$+U$\cite{Anisimov1997} study of the effects of altering the rare-earth in infinite-layer nickelates also performed a tentative comparison to a pressure-induced volume change without changing the chemical composition. The authors found that applying pressure results in a small increase of the hopping $t$ (estimated from the $d_{x^2-y^2}$ bandwidth), although they concluded that a substantially larger effect is found by the substitution of the rare-earth.

Using an effective single-orbital description of the nickelates, Kitatani {\it et al.} \cite{Kitatani2020} calculated $T_c$ of Sr-doped NdNiO$_2$ by means of the dynamical vertex approximation\cite{Rohringer2018} (D$\Gamma$A), predicting a superconducting dome which is in remarkably good agreement with that later found in experiments.\cite{Li2020,Zeng2020}
They furthermore argued that a decrease of the interaction-to-bandwidth ratio in their calculations would lead to an increase in $T_c$. It is therefore of interest to study the effect of pressure on the hopping $t$ and effective interaction $U$ to ascertain if such an effect would be consistent with the application of physical pressure in the nickelates.

In this work we use the recently developed multisite extension \cite{Petocchi2020b} of the multitier $GW$+EDMFT method \cite{Biermann2003,Ayral2013,Boehnke2016,Nilsson2017} to investigate the effects of pressure on the electronic properties of the infinite-layer nickelates. To emulate the effects of applying pressure, we assume the in-plane lattice constant to be fixed, and decrease only the out-of-plane lattice constant  $c$ (as speculated in Ref.~\onlinecite{Wang2022}).
We perform self-consistent $GW$+EDMFT simulations both for a close-to-optimally doped ($x=0.2$) and undoped ($x=0$) Pr$_{1-x}$Sr$_x$NiO$_2$ system. To accurately capture possible multi-orbital effects we use a low-energy model, obtained from downfolding an initial DFT\cite{Hohenberg1964,Kohn1965} calculation, containing the full Ni $3d$ manifold and the Pr 5$d_{xy}$ and 5$d_{z^2}$ orbitals.
We systematically study the effect of increasing physical pressure for both dopings, and find qualitative differences between the two pressure responses.
We show that a multi-band picture is important for describing the local state fluctuations in both the undoped and doped systems.
Intriguingly, however, the interplay between doping and pressure leads to an increasingly more single-band-like picture for the undoped system at high pressure, while this is not the case at (close-to) optimal doping. Here we instead find the multi-orbital nature to be important to describe the evolution of the electronic properties over the considered pressure range.

The paper is organized as follows. In Sec.~\ref{sec:Method} we describe the DFT calculations used as the starting point for the many-body calculations, followed by an outline of the multisite extension to the multitier $GW$+EDMFT method used in this work. In Sec.~\ref{sec:Results} we consider the effect of pressure on the electronic structure of the undoped and (close-to) optimally doped systems, and discuss what it implies for the single- versus multi-orbital nature of the materials. In Sec.~\ref{sec:Summary} we summarize our findings and conclusions. 

\section{\label{sec:Method}Method}

\subsection{DFT and technical details}

In our approach we start from a DFT\cite{Hohenberg1964,Kohn1965} calculation of Pr$_{1-x}$Sr$_x$NiO$_2$ in the infinite layered phase (space group $P4/mmm$). We use the generalized gradient approximation (GGA) \cite{Perdew1996} as implemented in the full-potential linearized augmented plane-wave (FLAPW)
code FLEUR \cite{Fleurcode} on a $16\times 16\times 16$ ${\bf k}$-point grid.
The Sr-doping is simulated using the Virtual Crystal Approximation\cite{Bellaiche2000} (VCA), and due to a technical limitation in the FLAPW method a fraction of the Pr is replaced by the consecutive element in the series, keeping the valence electrons consistent with a Sr doping. To discern possible differences in the pressure effects related to the doping, we will consider the pressure evolution of both the close-to-optimally doped ($x=0.2$) and the undoped ($x=0$) compound.

\begin{figure}[t]
 \includegraphics[width=0.49\textwidth]{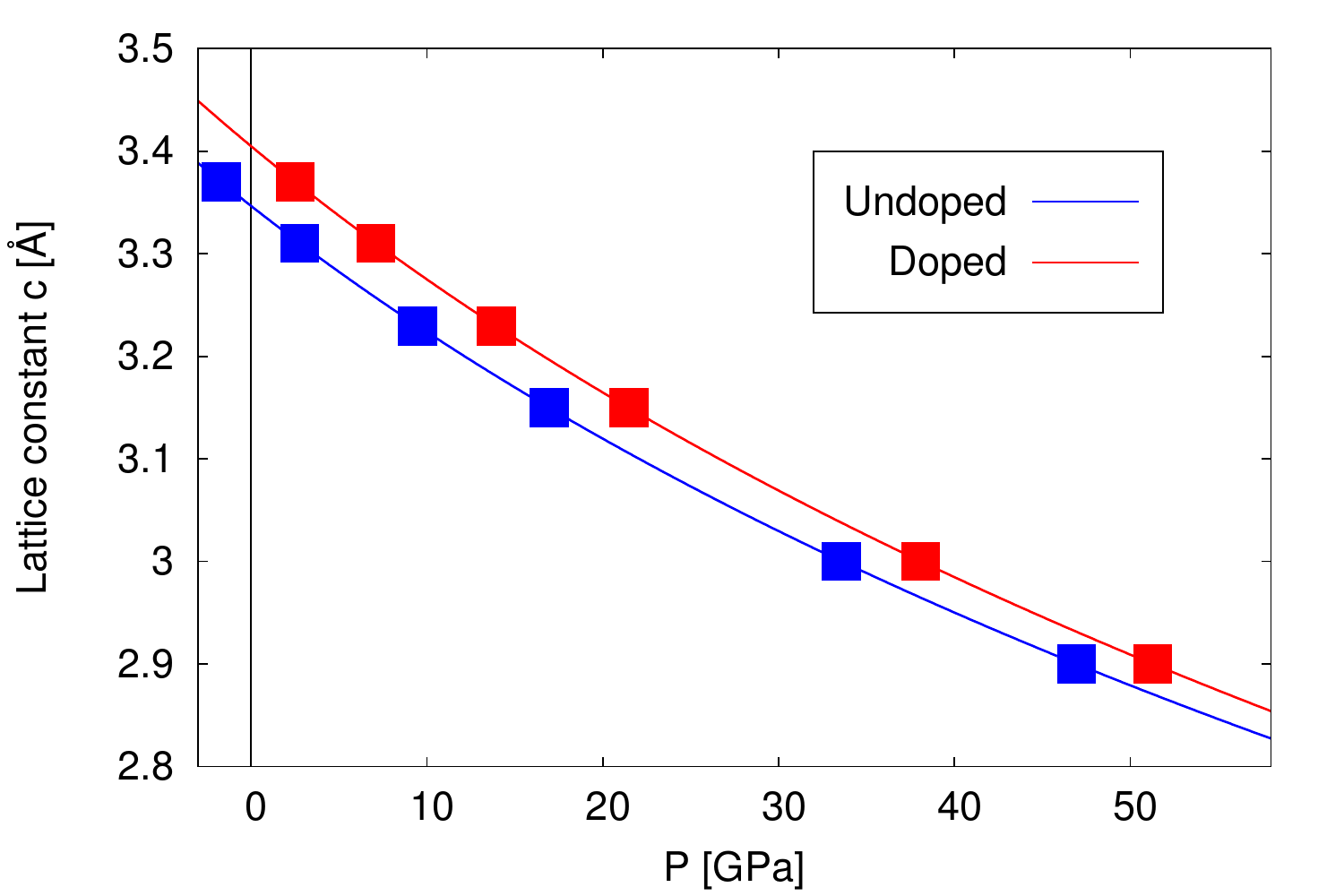}
  \caption{\label{Fig:PV} Lattice constant $c$ in \AA~as a function of pressure for undoped and doped ($x=0.2$) Pr$_{1-x}$Sr$_x$NiO$_2$. The squares indicate the lattice constants used for the calculations in this work.}
\end{figure}

The effect of increasing pressure is simulated by assuming the in-plane lattice constant $a=b=3.91$ \AA~to be kept fixed, 
similar to the approach in Ref.~\onlinecite{Been2021},\footnote{In this work, they also varied the in-plane lattice constant on the order of 1\% to follow more closely the effect of the change in the rare-earth element.}
and reduce only the out-of-plane lattice constant $c$. We give in Fig.~\ref{Fig:PV} an estimate of the theoretical increase in pressure corresponding to the decrease in lattice constant obtained by fitting the DFT results with the Vinet equation of state.\cite{Vinet1986}
The small offset of $\sim 5$ GPa between the two doping curves reflects the difference in the experimental lattice constants at zero pressure, which are reported to be $c=3.31$~\AA~for the undoped compound and $c=3.37$~\AA~at $x=0.2$ doping.\cite{Osada2020b}

The smallest lattice constant considered, $c=2.90$~\AA, corresponds to a 12\% (14\%) reduction for the undoped (doped) compound, respectively, and is consistent with the $\sim$10\% decrease found in $R$NiO$_2$ across the Lanthanides series,\cite{Been2021} confirming that our calculations in this pressure range are physically reasonable.
On the other hand, the largest lattice constant we have considered in this work, $c=3.37 $ \AA, is larger than the experimental value for the undoped compound at zero pressure. Accordingly it is estimated to correspond to a ``negative" theoretical pressure.

To treat the Pr $4f$ electrons we use, similarly to our previous work,\cite{Petocchi2020b} a manual core setup, and place the $4f^3$ electrons in the core. 
The remaining $4f$ states are treated with a self-consistent LDA~+~$U$~+~cRPA (constrained random-phase approximation\cite{Aryasetiawan2004}) scheme, where a cRPA calculation is performed iteratively to obtain a new interaction for the $f$-electrons in a LDA~+~$U$ calculation until convergence. Due to the observation of superconductivity in nickelates with different rare-earths,
it has been argued that these states are not very relevant for the pairing mechanism.\cite{Nomura2022} 
We therefore believe this treatment should not affect our predictions.

\begin{figure*}[t]
\centering
 \includegraphics[width=0.95\textwidth]{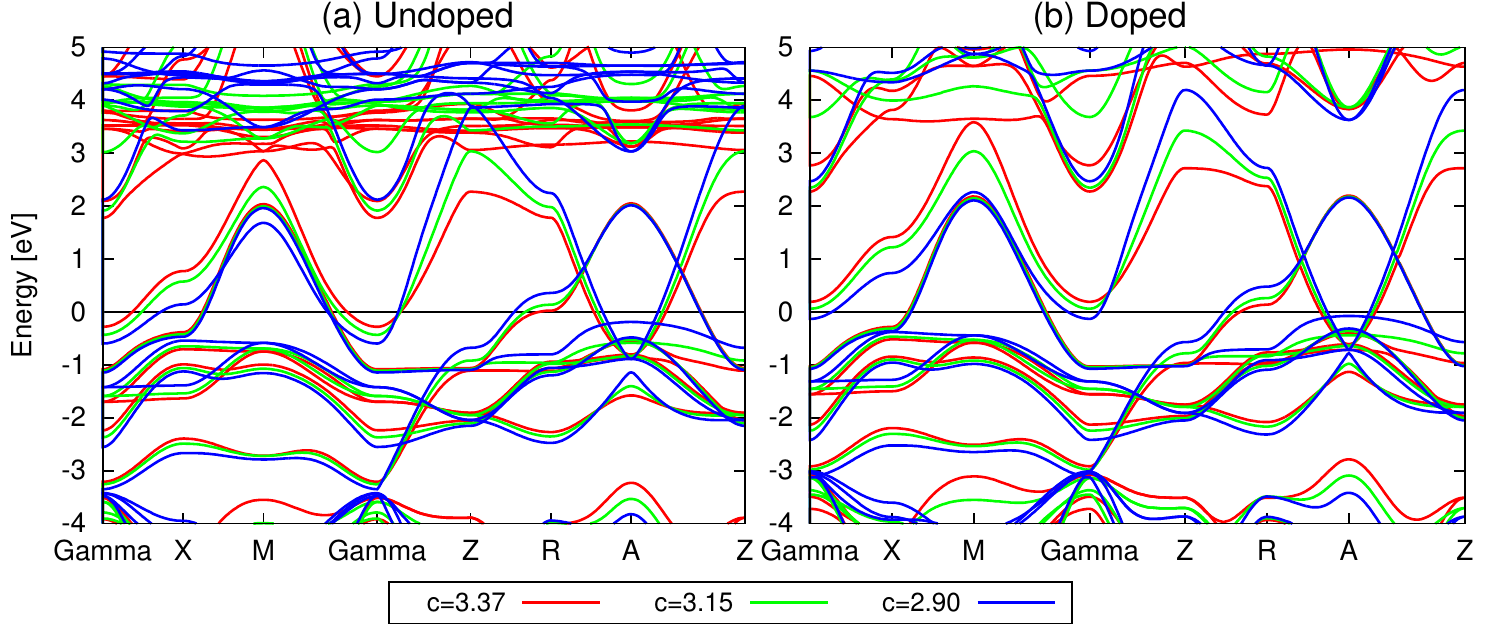}
  \caption{\label{Fig:BS} The DFT band structure for the three indicated lattice constants $c$ (for the equivalent pressure see Fig.~\ref{Fig:PV}) in (a) the undoped and (b) the doped compound.}
\end{figure*}
\begin{figure*}[t]
\centering
 \includegraphics[width=\textwidth]{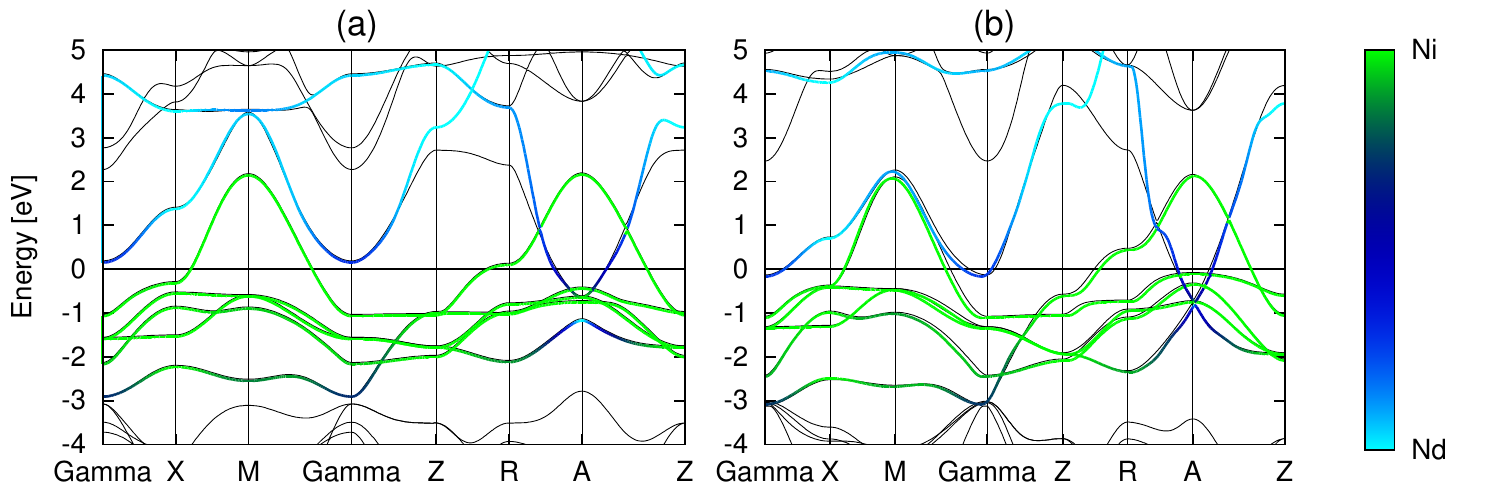}
  \caption{\label{Fig:BS Fatband} DFT and Wannier band structure with the relative contribution from the Ni and Nd orbitals to the bands indicated for the doped system at (a) zero pressure ($c=3.37$ \AA) and (b) the maximum pressure considered in this work ($c=2.90$ \AA).}
\end{figure*}

The DFT band structure for both the undoped and doped compounds are shown in Fig.~\ref{Fig:BS} as a function of decreasing lattice constant (increasing pressure). Interestingly, the self-doping Nd Fermi pocket around the $\Gamma$-momentum, which empties with increased Sr-doping $x$ and disappears near optimal doping, is again partially reappearing upon increasing pressure in the doped compound, as more clearly shown in Fig.~\ref{Fig:BS Fatband}.
The Nd pocket around the $A$-point instead slightly shrinks with increasing pressure. 
These DFT results are a first indication that chemical doping and physical pressure produce opposing effects. This phenomenon will become even more apparent in the discussions on the fully interacting system.

\subsection{$GW$+EDMFT}

In this section, we briefly outline the parameter-free multitier $GW$+EDMFT method and its recent multisite extension 
which has proven successful in the description of strongly correlated systems.\cite{Boehnke2016,Nilsson2017,Petocchi2020b,Petocchi2020a,Petocchi2021,Christiansson2022}
 For a more detailed discussion on the method, we refer to  Refs.~\onlinecite{Nilsson2017,Petocchi2020b}.

Starting from the Kohn-Sham eigenvalues and eigenfunctions obtained from the DFT calculation described in the previous section, we define a low-energy model using maximally-localized Wannier functions.\cite{Marzari1997,Mostofi2008} We adopt a 7-orbital model consisting of the five Ni $d$ orbitals, and of Nd 5$d_{xy}$ and 5$d_{z^2}$.
We then downfold the full DFT band structure to this low-energy subspace using a cRPA calculation \cite{Aryasetiawan2004} to obtain the effective bare interaction, $U_{\bf q}^\textrm{cRPA}(\omega)$, and a one-shot $GW$ \cite{Hedin1965} calculation ($G^0W^0$) to obtain the non-interacting propagator in the low-energy space, $G^{0}_{{\bf k}}$. The cRPA and $G^0W^0$ calculations were performed using the SPEX\cite{Friedrich2010} code with a $8\times 8\times 8$ ${\bf k}$-point grid and bands up to $\sim100$ eV used in the calculation of both the polarization and self-energy.

Within the multitier $GW$+EDMFT formalism,\cite{Nilsson2017} we solve the problem self-consistently using the extended dynamical mean field theory\cite{Georges1996,Sun2002} (EDMFT) self-consistency conditions $G^\textrm{loc}=G^\textrm{imp}$ and $W^\textrm{loc}=W^\textrm{imp}$, i.e., the local part of the Green's function and screened interaction equal the corresponding results 
obtained from the solution of the EDMFT impurity problem. 
The interacting lattice Green's function, with the contributions from the different tiers, takes the form
\begin{align}
\label{Eq:Gint}
G^{-1}_{{\bf k}}=&\,  i\omega_n + \mu -\varepsilon_{{\bf k}}^{\textrm{DFT}} + V^\text{XC}_{{\bf k}}  - \left(\Sigma_{{\bf k}}^{G^0W^0} - \Sigma_{{\bf k}}^{G^0W^0}\big|_{C} \right) \nonumber \\
& - \left( \Sigma_{{\bf k}}^{\textrm{sc}GW}\big|_{C} - \Sigma_{\mathrm{loc}}^{\textrm{sc}GW}\big|_{C} \right)-\Sigma_{\mathrm{loc}}^\textrm{EDMFT}\big|_{C},
\end{align}
where $\mu$ is the chemical potential, and the DFT exchange-correlation potential $V^\text{XC}_{{\bf k}}$, contained in the Kohn-Sham single-particle energies $\varepsilon_{{\bf k}}^{\textrm{DFT}}$, has been replaced by the $G^0W^0$ self-energy $\Sigma_{{\bf k}}^{G^0W^0}$.
The label $C$ denotes the correlated space encompassing all seven orbitals considered in the low-energy model. 
The double counting between the tiers is well-defined,\cite{Nilsson2017} and the $G^0W^0$ contribution coming from the states within the low-energy space, $\Sigma_{{\bf k}}^{G^0W^0}\big|_C$, is replaced by the self-consistently obtained $GW$ self-energy $\Sigma_{{\bf k}}^{\textrm{sc}GW}\big|_C$. An increased level of accuracy in the treatment of strong correlations is then achieved by replacing its local projection, $\Sigma_{\mathrm{loc}}^{\textrm{sc}GW}\big|_C$, by the EDMFT impurity self-energy, $\Sigma_{\mathrm{loc}}^\textrm{EDMFT}\big|_C$. 

The contributions to the screening of the interaction from the different tiers and the corresponding double counting terms are similarly obtained,
\begin{align}
W^{-1}_{{\bf k}}=& v_{{\bf k}}^{-1}- \left(\Pi_{{\bf k}}^{G^0G^0} - \Pi_{{\bf k}}^{G^0G^0}\big|_{C} \right) \nonumber \\
& - \left( \Pi_{{\bf k}}^{GG}\big|_{C} - \Pi_{\mathrm{loc}}^{GG}\big|_{C} \right  )-\Pi_{\mathrm{loc}}^\textrm{EDMFT}\big|_{C},
\end{align}\\
with the bare interaction $v_{{\bf k}}$ screened by the polarization contributions from the RPA ($\Pi_{{\bf k}}^{G^0G^0}$), self-consistent $GW$ ($\Pi_{{\bf k}}^{GG}$), and the impurity $\Pi_{\mathrm{loc}}^\textrm{EDMFT}$.

To handle the Ni and Nd sites, the multisite extension of $GW$+EDMFT \cite{Petocchi2020b} defines separate fermionic and bosonic Weiss fields according to
\begin{align}
& \mathcal{G}_i = \left( \Sigma^\textrm{imp}_i+(G_i^\textrm{imp})^{-1}  \right)^{-1} ,\\
& \mathcal{U}_i = W^\textrm{imp}_i \left(1+\Pi^\textrm{imp}_i W^\textrm{imp}_i \right)^{-1} ,
\end{align}
with the indices $i\in$\{Ni, Nd\}.
This results in two separate impurity problems which are solved using a continuous-time Monte Carlo solver \cite{Werner2006,Hafermann2013} capable of treating dynamically screened interactions.\cite{Werner2010}
The resulting $\Sigma_i^\textrm{imp}$ and $\Pi_i^\textrm{imp}$ for the two impurities are subsequently coupled in the lattice self-consistency 
equations, where the EDMFT self-consistency conditions 
\begin{align}
G_i^\textrm{imp} &= G_i^\textrm{loc},  \quad
W_i^\textrm{imp} = W_i^\textrm{loc} 
\end{align}
now have to be fulfilled for each of the two sites $i\in$\{Ni, Nd\}.

\section{\label{sec:Results}Results}

The $GW$+EDMFT calculations were performed at temperature $T=1/30$ eV on the $8\times 8\times 8$ ${\bf k}$-point grid from the downfolding, using the low-energy space defined in the previous section. The presented results have been obtained from an average over at least 10 consecutive converged iterations. We remind the reader that the $c=3.37$ \AA~lattice constant is larger than the experimental value and hence corresponds to a theoretical negative pressure (Fig.~\ref{Fig:PV}). The corresponding data points are included for completeness, but left out of the discussions.

\subsection{Orbital occupations}

\begin{figure}[t]
 \includegraphics[width=0.49\textwidth]{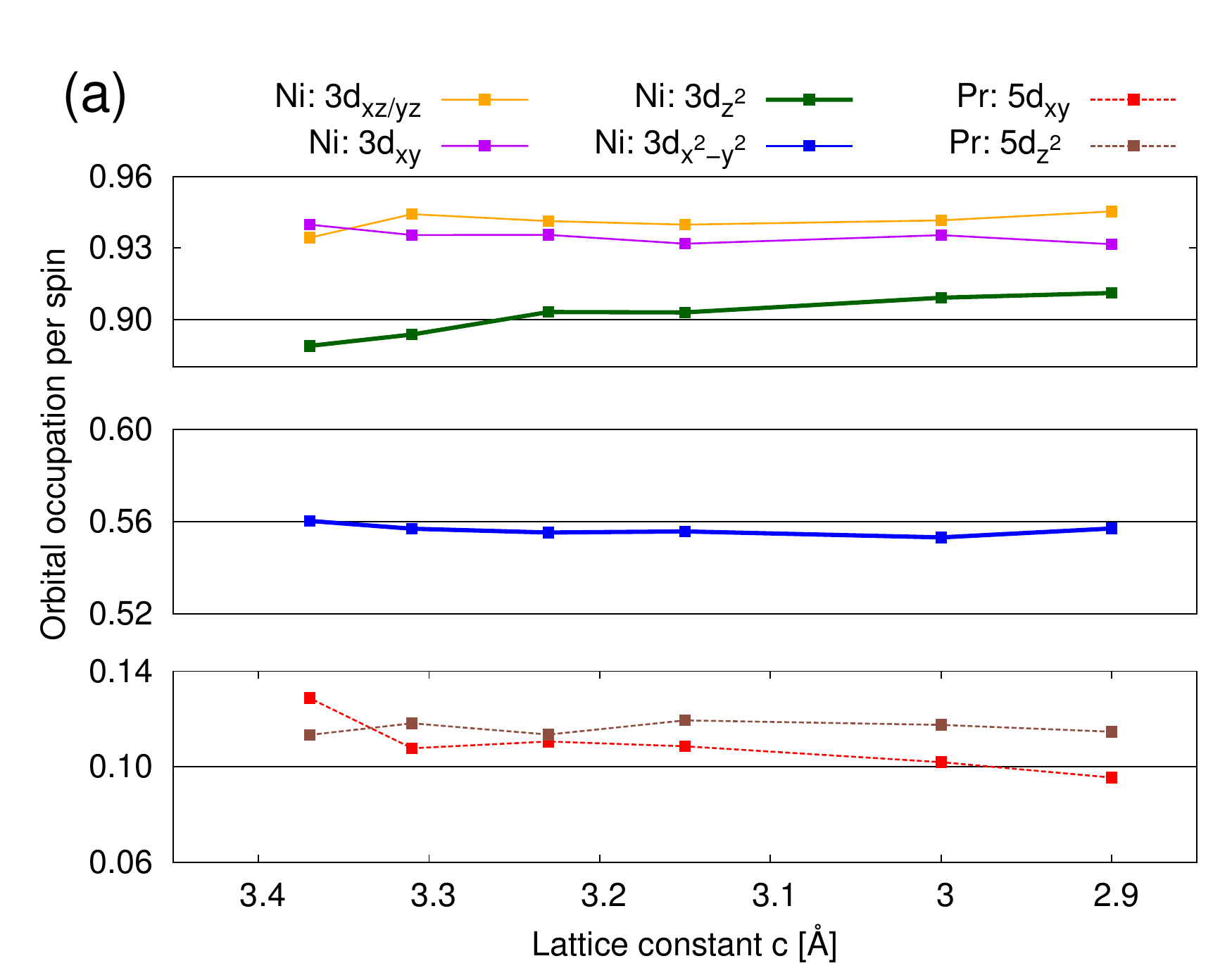}
  \includegraphics[width=0.49\textwidth]{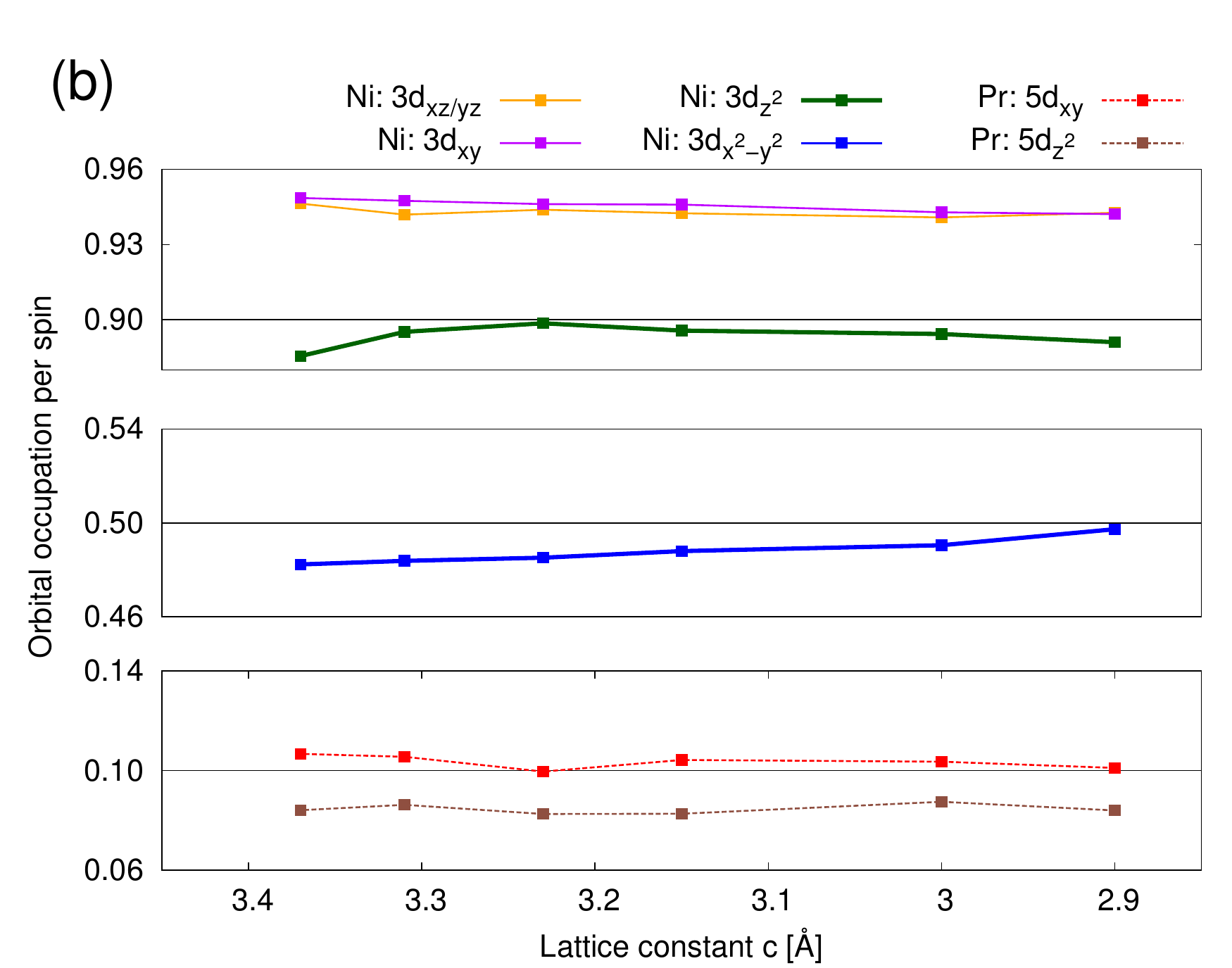}
  \caption{\label{Fig:Orbital occ}  Pressure dependence (in terms of the lattice constant $c$) of the orbital-resolved occupation per spin of the five Ni and the two Pr $d$-like orbitals included in the model for (a) the undoped and (b) the doped systems. The error bars given by the standard deviation for the averaged iterations is smaller than the dot sizes.
Note the different ranges in the middle panels for the Ni 3$d_{x^2-y^2}$ (blue) orbitals.
  }
\end{figure}

\begin{figure}[t]
 \includegraphics[width=0.49\textwidth]{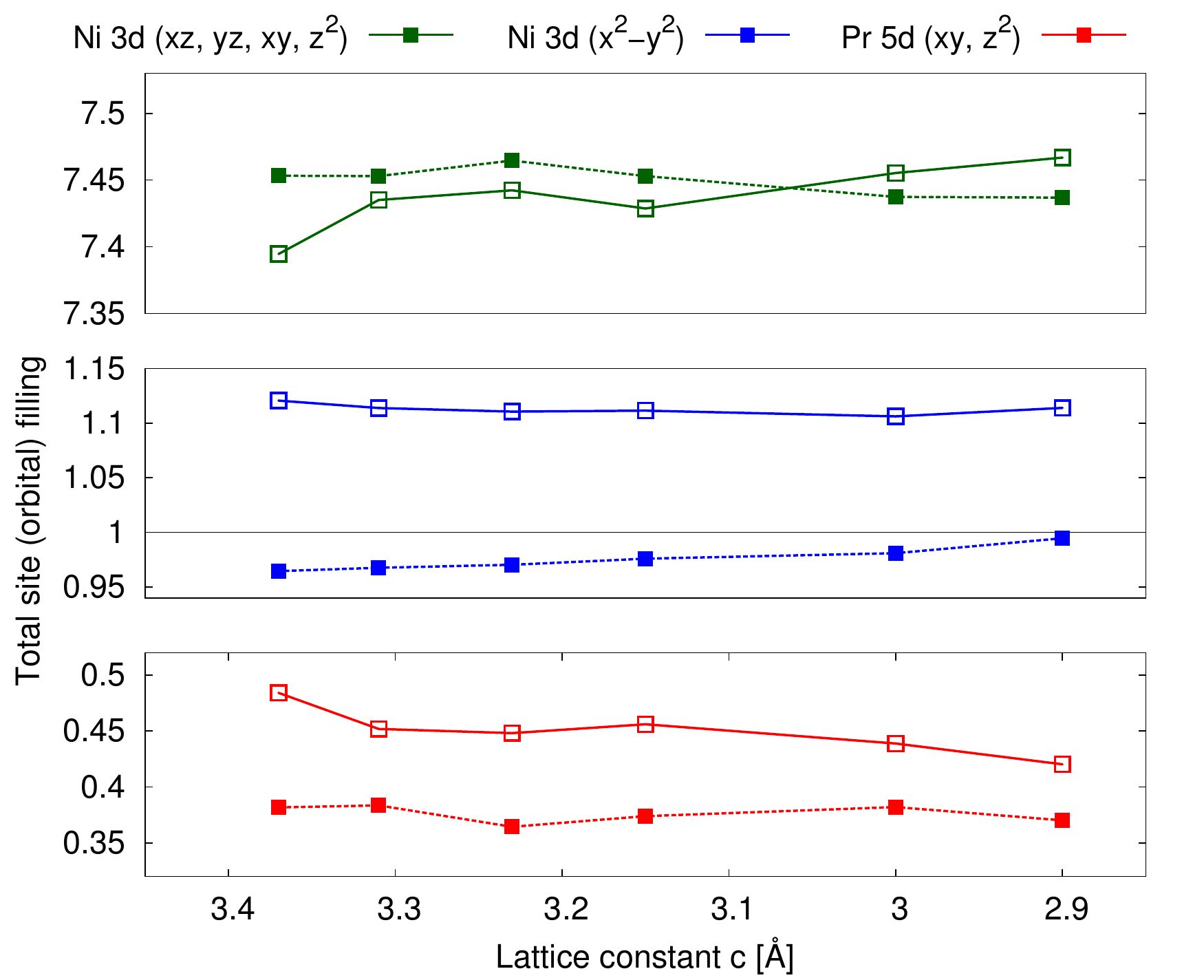}
  \caption{\label{Fig:Site Occ}  Pressure dependence of the filling of the Ni and Pr sites (orbitals), in terms of the lattice constant $c$. We plot the total charge in the orbitals listed in the labels. The Ni 3$d_{x^2-y^2}$ (close to half-filling) is shown separately from the other orbitals on the Ni site to clarify where the charge is moving. Full lines with empty squares show the results for the undoped compound and the dashed lines with filled squares show the results for the doped compound. }
 \end{figure}

In Fig.~\ref{Fig:Orbital occ} we show the pressure dependence of the orbital occupations in the undoped and doped compounds. From these orbital resolved occupations, we can deduce two main trends. In the undoped system, the Ni 3$d_{x^2-y^2}$ filling remains pinned around a constant occupation of 0.56 electrons/spin, slightly higher than half-filling, and the main change is found in the Ni 3$d_{z^2}$-like orbital, whose filling monotonically increases with pressure. To better understand the charge redistribution, we look also at the site resolved occupations in Fig.~\ref{Fig:Site Occ}, where the interesting Ni 3$d_{x^2-y^2}$ orbital close to half-filling is shown separately.
From this, it becomes clear that with increasing pressure, charge from the Pr site is transferred to the almost filled Ni orbitals, and that in particular the less occupied Ni 3$d_{z^2}$ orbital is filled. 
The effect of increasing pressure is hence to suppress the self-doping, potentially making the system more single-band like.
 
The above trends in the undoped compound are in stark contrast to the behavior observed in the doped compound. Already the orbital resolved occupations indicate a transfer of charge away from the almost filled Ni orbitals to the Ni 3$d_{x^2-y^2}$ orbital, shifting it closer to half-filling, while the occupation of the Pr site is only slightly decreasing with pressure, as is visible in Fig.~\ref{Fig:Site Occ}.
Furthermore, in the doped compound, the filling of the Ni 3$d_{z^2}$ orbital displays a non-trivial pressure-dependence -- at low pressures the occupation is increasing, while at higher pressures it empties out again, leading to the increase of the Ni 3$d_{x^2-y^2}$ occupation towards half-filling at high pressure.
This different behavior in the orbital occupations is a first indication of the importance of considering the presence of additional Ni orbitals to accurately capture the pressure dependence
of the doped system, while this is less evident in the case of the undoped compound.

We note that unlike in our previous study of NdNiO$_2$,\cite{Petocchi2020b} (at the higher temperature $T=1/10$ eV) we now find indications for a tendency to magnetic ordering (changes in the orientation of the magnetic moment between iterations, but not yet a stable order)
at $T=1/30$ eV. This tendency towards magnetic ordering is significantly larger in the doped compound, compared to the undoped system.
However, since we performed the calculations for this study using the primitive cell, a more detailed analysis of the magnetic ordering is beyond the scope of this work, and we refer the reader to Ref.~\onlinecite{Nomura2022} and references therein.

\subsection{Interaction parameters}

A powerful feature of the $GW$+EDMFT formalism is that it provides a self-consistent calculation of the effective interaction parameters, which takes into account the screening by charge fluctuations in the low-energy subspace. 
In this section, we discuss the pressure-induced changes in these interaction parameters, and how they are affected by doping. We focus on the effective bare on-site interaction, $\mathcal{U}(i\omega_n=0)$, and the Hund coupling, $\mathcal{J}(i\omega_n=0)$. 

In the undoped case (Table~\ref{Table:Undoped}), the mostly occupied Ni orbitals (Ni 3$d_{xz/yz}$, $d_{xy}$, $d_{z^2}$) show a small decrease of the interaction $\mathcal{U}$ in the high-pressure regime, the most notable change being observed for the $3d_{z^2}$ orbital with a decrease of around 0.14 eV over the considered pressure range. The $d_{x^2-y^2}$-like orbital (close to half-filling) instead exhibits a minor increase of $\sim0.07$~eV in the interaction strength. 

 \begin{table*}[t]
\setlength{\tabcolsep}{15pt} 
\renewcommand{\arraystretch}{1.2}
\caption{\label{Table:Undoped} Pressure dependence of the diagonal elements of the effective interaction $\mathcal{U}(i\omega_n=0)$ in eV for the undoped system.}
 \begin{tabular}{|c||c c c c c c|}
 \hline
 	Orbital & 3.37 & 3.31 & 3.23 & 3.15 & 3.00 & 2.90 \\
 \hline
 \hline 
  Ni 3$d_{xz/yz}$ & 5.53 & 5.49 & 5.47 & 5.53 & 5.48 & 5.47 \\
  Ni 3$d_{xy}$ & 5.27 & 5.23 & 5.21 & 5.24 & 5.17 & 5.14  \\
  Ni 3$d_{z^2}$ & 5.46 & 5.37 & 5.33 & 5.38  & 5.27 & 5.23  \\
  Ni 3$d_{x^2-y^2}$ & 4.52 & 4.48 & 4.47 & 4.54  & 4.52 & 4.55  \\
\hline
\hline
  Pr 5$d_{xy}$ & 1.95 & 1.99 & 1.98 & 1.92 & 1.93 & 1.90  \\
  Pr 5$d_{z^2}$ & 1.84  & 1.88 & 1.95 & 1.87 & 1.94 & 1.91 \\
\hline
 \end{tabular}
 \end{table*}
  \begin{table*}[t]
\setlength{\tabcolsep}{15pt}
\renewcommand{\arraystretch}{1.2}
\caption{\label{Table:Doped}Pressure dependence of the diagonal elements of the effective interaction $\mathcal{U}(i\omega_n=0)$ in eV for the doped system.}
 \begin{tabular}{|c||c c c c c c|}
 \hline
 	Orbital & 3.37 & 3.31 & 3.23 & 3.15 & 3.00 & 2.90 \\
 \hline
 \hline 
  Ni 3$d_{xz/yz}$ & 5.66 & 5.66 & 5.65 & 5.66 & 5.68 & 5.67 \\
  Ni 3$d_{xy}$ & 5.18 & 5.19 & 5.20 & 5.20 & 5.20 & 5.16 \\
  Ni 3$d_{z^2}$ & 5.79 & 5.75 & 5.70 & 5.68 & 5.82 & 5.79 \\
  Ni 3$d_{x^2-y^2}$ & 4.43 & 4.46 & 4.49 & 4.51 & 4.59  & 4.63  \\
\hline
\hline
  Pr 5$d_{xy}$ & 2.05  & 2.04 & 2.05 & 2.02 & 1.82 & 1.84  \\
  Pr 5$d_{z^2}$ & 1.96 & 2.02 & 2.05 & 2.04 & 1.82 & 1.77 \\
\hline
 \end{tabular}
 \end{table*}

The situation is again very different for the doped system (Table~\ref{Table:Doped}); the $\mathcal{U}$ for the Ni 3$d_{xz/yz}$ and $d_{xy}$ orbitals remain approximately constant with only very small fluctuations with pressure, while the $3d_{z^2}$ orbital shows a decrease followed by an abrupt increase of around 0.14 eV at high pressures, where the orbital occupation is again decreasing (compare to Fig.~\ref{Fig:Orbital occ}). Such a non-monotonic pressure effect on the effective interaction has previously been discussed in the context of cRPA,\cite{Tomczak2009} where the authors argued that the origin is a competing effect on the polarization from the band structure change and the orbital overlaps. These two effects tend to decrease and increase the polarization under pressure, respectively.

The $d_{x^2-y^2}$ orbital, on the other hand, displays a clear increase of 0.20 eV over the considered pressure range, almost three times the increase found in the undoped system, and again following the orbital occupation. This leads to the unexpected\cite{Kitatani2020,Held2022} 
result of an effective interaction $\mathcal{U}$ which in the doped system is {\it larger} at high pressure than at ambient conditions.

On the almost empty Pr sites, in both the undoped and doped systems, the interaction on the 5$d_{xy}$ orbitals is lowered with increasing pressure, by 0.09 and 0.20 eV respectively. While the undoped compound displays a monotonic behavior, the doped system again exhibits an abrupt change in the high-pressure region. Similarly, the interaction for the 5$d_{z^2}$ orbital in the undoped case fluctuates, while the doped system initially sees a slight increase with pressure followed by a rapid decrease of around 10\%, without any discernible corresponding change in the orbital occupation (see Fig.~\ref{Fig:Orbital occ}).

We also note that $U^\text{cRPA}$ for the important Ni 3$d_{x^2-y^2}$ orbital follows approximately the same trend as discussed above for $\mathcal{U}$, but overestimates the interaction strengths (since it lacks the nonlocal screening from the low-energy subspace). For the undoped system at the experimental lattice parameters we have $U_u^\text{cRPA}=4.97$ and for the doped system $U_d^\text{cRPA}=5.00$ eV, compared to $\mathcal{U}_u=4.48$ and $\mathcal{U}_d=4.43$ eV. 
It is worth mentioning also the overestimation of $U^\text{cRPA}$ for the Pr orbital, which is almost 40\% larger than the self-consistently calculated local effective bare interaction $\mathcal{U}$ (2.61 eV compared to 1.88 eV for the undoped, and 2.68 eV compared to 1.96 eV for the doped compound).
These considerations emphasize again the importance of treating the low-energy physics beyond DFT and cRPA to accurately describe and compare the doped and undoped systems.

\begin{figure}[t]
\includegraphics[width=0.49\textwidth]{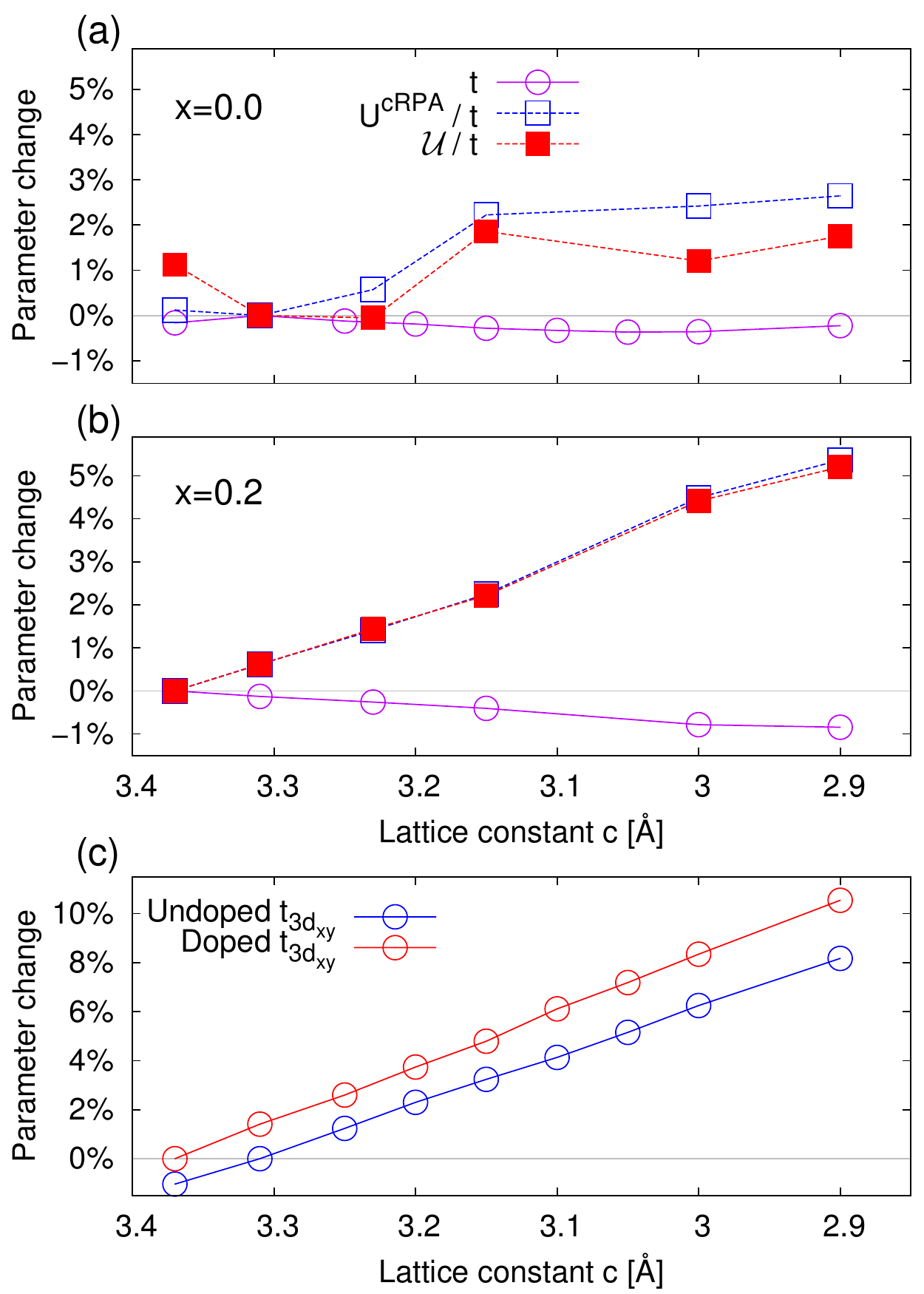}
 \caption{\label{Fig:Ut} Change in \% from the zero-pressure values of the effective hopping $t$ derived from the Wannier Hamiltonian, and the interaction-to-hopping ratios $U^{cRPA}/t$ and $\mathcal{U}/t$ for Ni $3d_{x^2-y^2}$ as a function of pressure in (a) the undoped and (b) the doped compound. (c) Change compared to the zero-pressure result in the hopping $t$ for Ni $3d_{xy}$ derived from the Wannier Hamiltonian. 
}
\end{figure}

In Ref.~\onlinecite{Kitatani2020} it was hypothesized that a smaller $U/t$ ratio under pressure could be responsible for raising $T_c$ based on their $D\Gamma A$ calculation for a single-orbital model. To elucidate this point, 
we compare the calculated in-plane hopping parameter for the Ni $3d_{x^2-y^2}$ in the Wannier Hamiltonian, $t=-\mathcal{H}\big({\bf R}=(1,0,0)\big)$, to the $U^\text{cRPA}$ and $\mathcal{U}$ interaction in Fig.~\ref{Fig:Ut}.

The hopping $t$ in the undoped system behaves qualitatively similar (although quantitatively different) to the Wannier estimate by Been {\it et al.},\cite{Been2021} who studied the dependence on the rare-earths and thereby effectively also changed the lattice parameter, with a non-monotonic change (decrease followed by a subsequent increase) in the hopping over the pressure range.
In the doped compound, we instead see a monotonic {\it decrease} in the hopping with pressure. 

Although it is known that the effective interaction can increase with applied pressure,\cite{Tomczak2009} the observed behavior of the hopping parameter is rather counter-intuitive, as one would naively expect it to increase due to a larger overlap of the orbitals. We can however understand this trend by looking at the evolution of the Ni $3d_{x^2-y^2}$ Wannier functions; contrary to expectation, they become more localized with pressure resulting in smaller hopping integrals.
Together with the previously discussed pressure dependence of the interaction, this leads to a significant difference in the doping behavior, with the undoped compound showing a more or less constant ratio of the interaction to hopping, $U/t$, while the doped compound exhibits a systematic {\it increase} with increasing pressure.  
All other Wannier functions instead exhibit the expected behavior of delocalization with increasing pressure, and as a consequence we observe for the other in-plane Ni $3d_{xy}$ orbital the expected increase in $t_{3d_{xy}}$, see Fig.~\ref{Fig:Ut} (c).

Another estimate of the hopping parameters in the undoped compound was given in Ref.~\onlinecite{Been2021} based on the band width of Ni $3d_{x^2-y^2}$ for NdNiO$_2$, which they found to increase by $\sim5$ \% over our studied range. Using a similar estimate, we only find a change of 1-2\%. This discrepancy can be traced back to the different treatment of the in-plane ($a=b$) lattice constant, which we have kept fixed. If we similarly also decrease the in-plane lattice constant by $\sim 1$\% we find an equivalent change of $\sim$4\% in the  Ni $3d_{x^2-y^2}$ bandwidth. Using instead this estimated change in $t$ for the undoped system would result in a slightly decreased ratio $U/t$ at high pressure, while we would still not observe a decrease in the doped system -- it would instead remain approximately constant (or slightly increase).
 With either way of estimating the hopping parameter, it becomes clear from the behavior of $U/t$ that the undoped and doped systems react in qualitatively different ways to pressure. This makes it difficult to provide a simple explanation for the experimentally observed increase in $T_c$ with increasing pressure based solely on this estimate.

 \begin{table*}[t]
\setlength{\tabcolsep}{15pt} 
\renewcommand{\arraystretch}{1.2} 
\caption{\label{Table:Hund J} Pressure dependence of the effective Hund coupling $\mathcal{J}_{ab}(i\omega_n=0)$ in eV between $a=\text{Ni 3}d_{x^2-y^2}$ and the other Ni orbitals  for the undoped (u) and doped (d) systems in eV.}
 \begin{tabular}{|r||c c c c c c|}
 \hline
 $b$ & 3.37 & 3.31 & 3.23 & 3.15 & 3.00 & 2.90 \\
 \hline
 \hline 
  Ni 3$d_{z^2}$ (u) & 0.65 & 0.65 & 0.65 & 0.65 & 0.64 & 0.64 \\
  Ni 3$d_{xy}$ (u) & 0.36 & 0.36 & 0.36 & 0.37 & 0.37 & 0.36  \\
  Ni 3$d_{xz/yz}$ (u) & 0.60 & 0.60 & 0.60 & 0.61 & 0.61 & 0.61 \\
 \hline
 \end{tabular}
 \begin{tabular}{|r||c c c c c c|}
 \hline
  Ni 3$d_{z^2}$ (d) & 0.65 & 0.65 & 0.65 & 0.65 & 0.65 & 0.65 \\
  Ni 3$d_{xy}$(d)  & 0.35 & 0.35 & 0.35 & 0.35 & 0.35 & 0.35   \\
  Ni 3$d_{xz/yz}$ (d) & 0.58 & 0.58 & 0.58 & 0.59 & 0.59 & 0.59  \\
\hline
 \end{tabular}
 \end{table*}

Finally we also briefly consider the Hund coupling $\mathcal{J}$. Similarly to NdNiO$_2$,\cite{Petocchi2020b} the calculated $\mathcal{J}$ are overall slightly decreased under hole-doping. We show the couplings involving the Ni $3d_{x^2-y^2}$-like orbital in Table~\ref{Table:Hund J}. Contrary to the intraorbital interactions $\mathcal{U}$, however, the values of $\mathcal{J}$ remain mostly unaffected by pressure, with maximum changes on the order of $\sim 0.01$ eV.

To wrap up the discussion on the interaction parameters we note that although the overall changes in the interaction and hopping parameters are small to moderate at most, we observe clear systematic trends. The results indicate a qualitatively different effect of pressure in the undoped and doped systems, in particular in the behavior of the interaction-to-hopping ratio, which must be taken into account when considering its implications for superconductivity.

\subsection{Statistics}

 \begin{figure}[t]
 \includegraphics[width=0.49\textwidth]{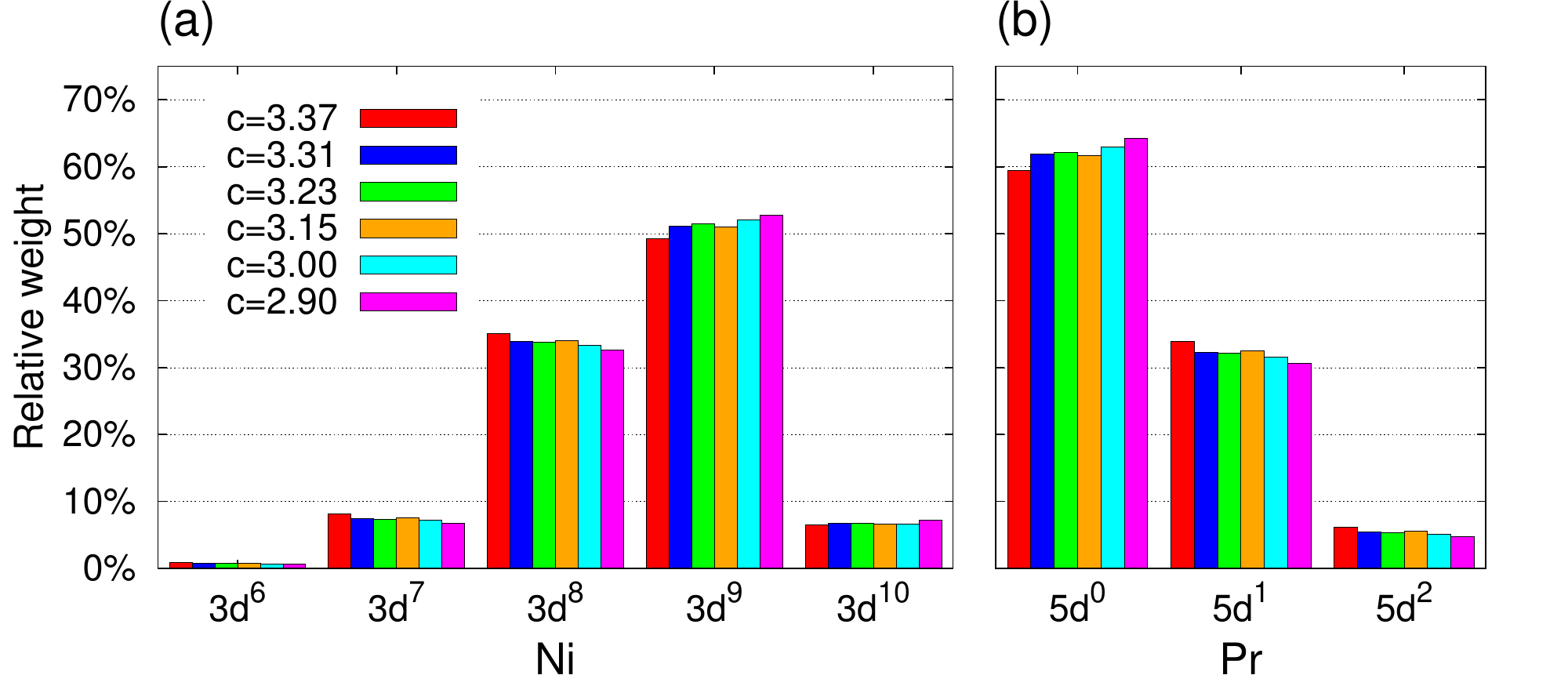}
 \includegraphics[width=0.49\textwidth]{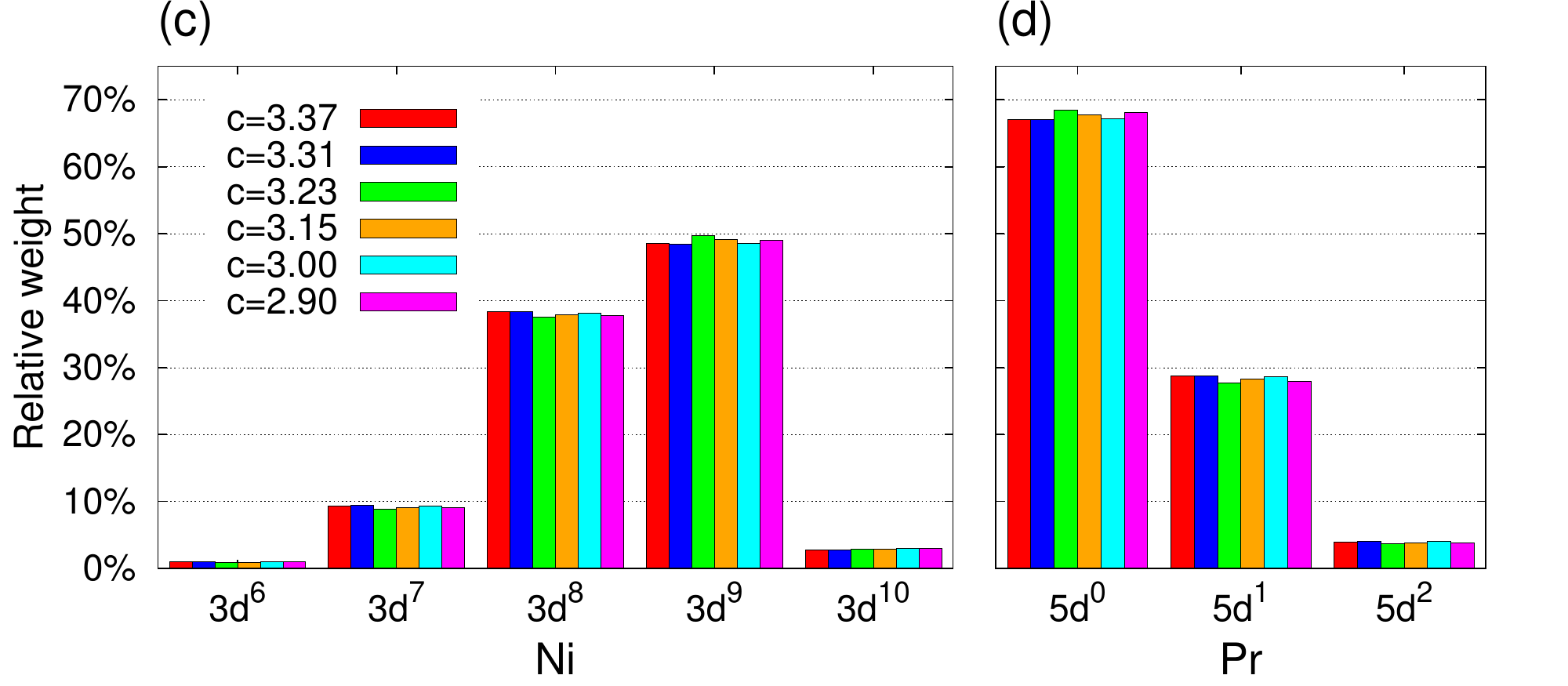}
  \caption{\label{Fig:Config Statistics} EDMFT configuration statistics in the undoped system for (a) Ni and (b) Pr, and the doped system for (c) Ni and (d) Pr. }
 \end{figure}
 \begin{figure}[t]
 \includegraphics[width=0.49\textwidth]{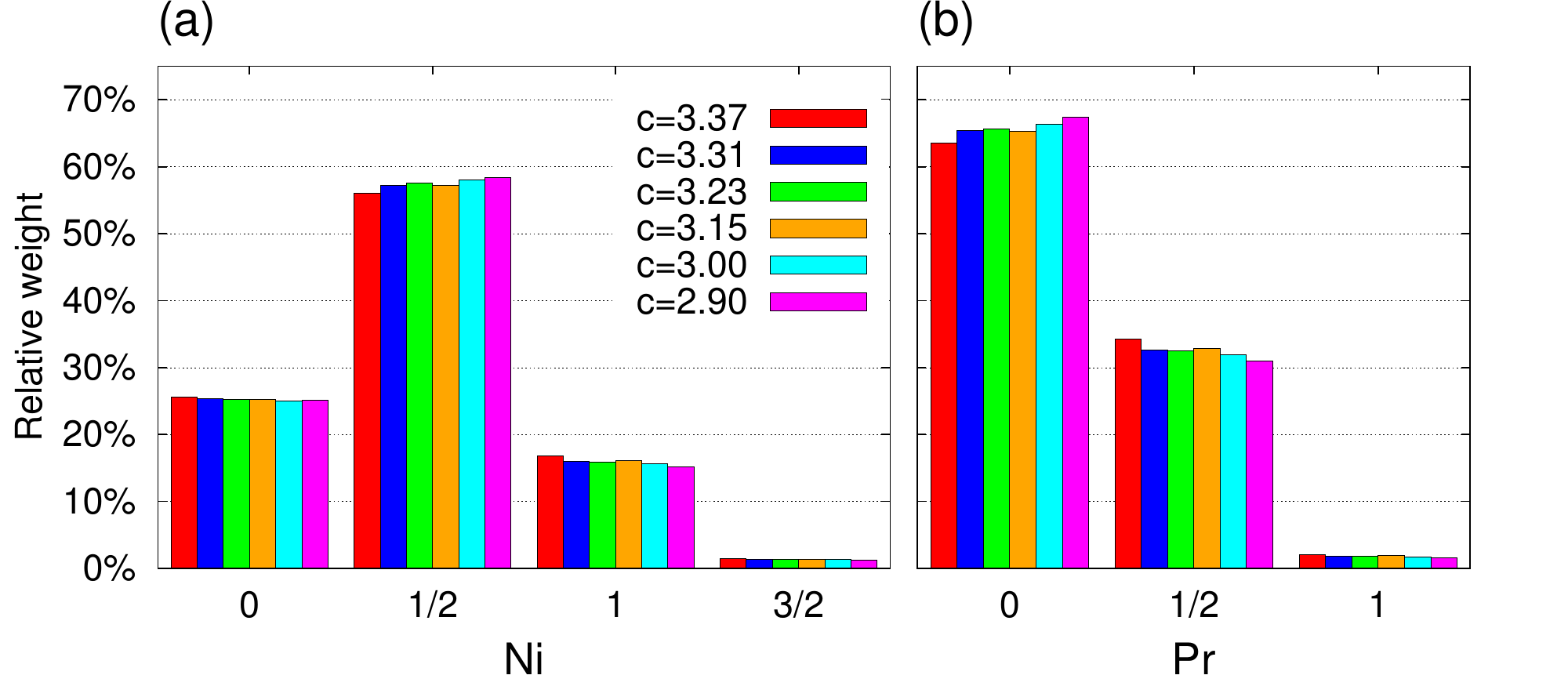}
 \includegraphics[width=0.49\textwidth]{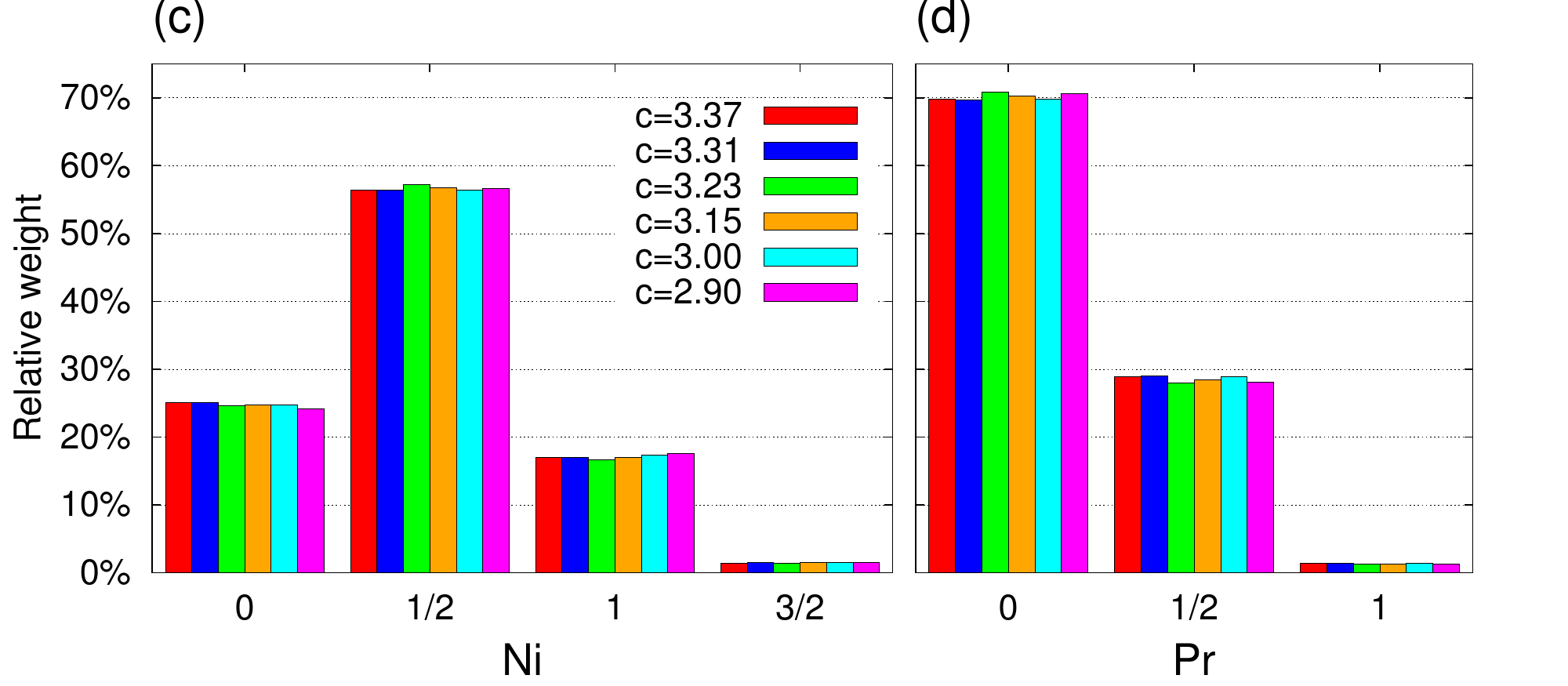}
  \caption{\label{Fig:Spin Statistics} EDMFT spin statistics for the spin-states $S_z$ in the undoped system for (a) Ni and (b) Pr, and the doped system for (c) Ni and (d) Pr.}
 \end{figure}

To further study the pressure evolution of PrNiO$_2$ and the single-versus-multi-orbital characteristics of the system, we next discuss the statistics of the populated charge and spin sectors, as obtained from the EDMFT calculations. By analyzing the corresponding histograms, we can gain insights into the relevant atomic states. We show in Fig.~\ref{Fig:Config Statistics} and \ref{Fig:Spin Statistics} the occupation and spin statistics, respectively, for the undoped and doped compounds.

In agreement with our previous calculations for NdNiO$_2$,\cite{Petocchi2020b}
the holes doped into the system empty the rare-earth site, as is evident from the increase in the Pr $5d^0$ configuration and corresponding decrease in the $5d^1$  configurations at fixed pressure (lattice constant).
Upon increasing pressure, the undoped system shows weak indications for an increasingly single-band-like situation with an increase in the Ni $3d^9$ and Pr $5d^0$ configurations, while the fluctuations to Ni $3d^8$ and Pr $5d^1$ are suppressed accordingly -- in agreement with the previous discussions of the orbital occupations. 
The effect of doping the system is an increase in the Ni $3d^7$ and $3d^8$ configurations, and a reduction in the $3d^9$ weights, in agreement with what would naively be expected. 
To interpret the pressure dependence and the trend towards a more or less single-band-like picture one needs to take into account the charge transfer between Pr and Ni. Here, we notice that in the doped compound, there is almost no pressure dependent change in the Pr $5d^0$ weight. Hence, in the doped compound the self-doping from Pr is essentially unaffected by pressure, in contrast to the undoped compound where we observe a clear reduction. This is consistent with our previous observation that the orbital occupations are mostly redistributed locally (on the same site) by pressure in the doped system. Given this fact and the almost pressure-independent Ni $3d^n$ occupations, we conclude that the occupation statistics of the doped compound shows no hint of a more single-band-like behavior with increasing pressure.

The histograms of the spin states, shown in Fig.~\ref{Fig:Spin Statistics}, indicate the importance of a multi-orbital description of both systems, because of the large weight from the high-spin configurations, as argued previously also for NdNiO$_2$.\cite{Petocchi2020b}  
Similarly to the charge statistics, the undoped compound shows the behavior expected for an increasingly single-band-like description with increasing pressure, with a slight increase of the $S_z^\textrm{Ni}=\frac{1}{2}$ and $S_z^\textrm{Pr}=0$ states, whereas the doped system again does not display any such systematic change.
 
 \begin{figure*}[t]
 \includegraphics[width=0.99\textwidth]{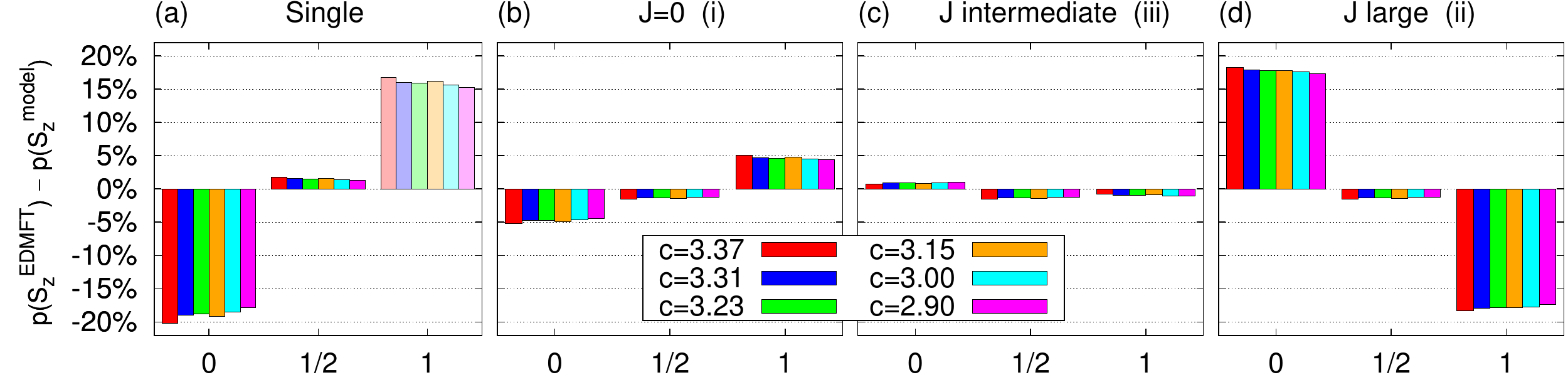}
 \includegraphics[width=0.99\textwidth]{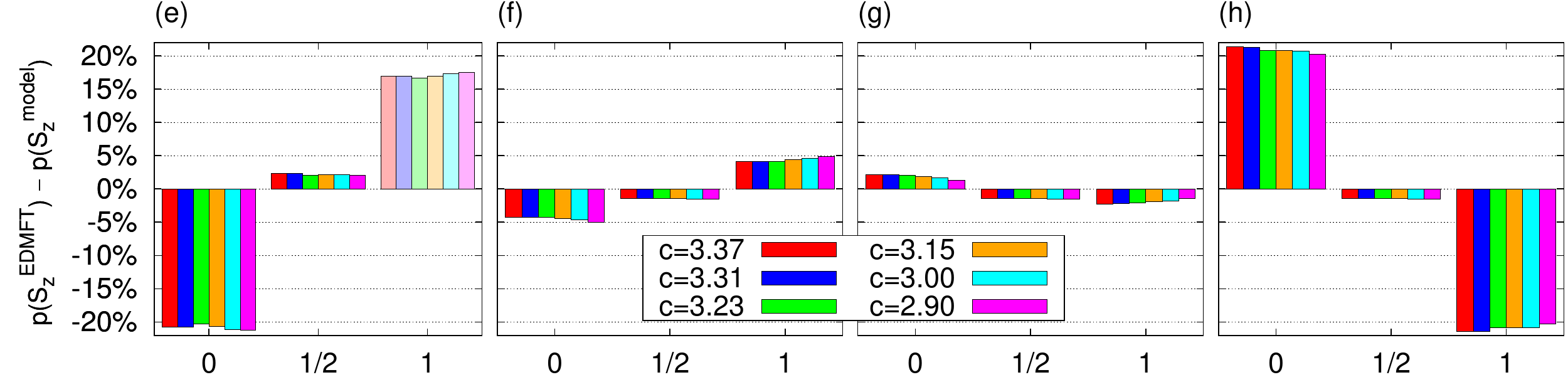}
  \caption{\label{Fig:Spin model} Difference in the model estimates for the spin-state ($S_z$=0, 1/2, 1) weights from the full EDMFT results ($p(S_z^\text{EDMFT})-p(S_z^\text{model})$) for the undoped (a)-(d) and the doped (e)-(h) systems within the single- and two-orbital models described in the text. Since the probability for $S_z=1$ in the one-orbital model is 0, the light shaded histograms in the left-most panels indicate the missing weight of the high-spin states.
  }
 \end{figure*}

\begin{table*}[t]
\caption{\label{Table:Model spin} Contributions to the spin states $S_z$ from the $n$-electron sectors with probabilities $p(3d^{n})$ for the models discussed in the text. The probabilities are taken from the EDMFT configuration statistics shown in Fig.~\ref{Fig:Config Statistics}.
}

\renewcommand{\arraystretch}{1.5}
\setlength{\tabcolsep}{8pt}
\begin{tabular}{|c|c|c|c|c|c|}
\hline
& $n=0$ & $n=1$ & $n=2$ & $n=3$ & $n=4$ \\
\hline\hline
One-orbital &  $S_z=0$ : $p(3d^8)$   %
		& $S_z=\frac{1}{2}$ : $p(3d^9)$ %
		& $S_z=0$ : $p(3d^{10})$ & --  & --  \\
\hline
Two-orbital (i) &  $S_z=0$ : $p(3d^6)$  %
		& $S_z=\frac{1}{2}$ : $p(3d^7)$ %
		& $S_z=0$ : $\frac{4}{6}p(3d^8)$  %
		& $S_z=\frac{1}{2}$ : $p(3d^9)$ %
		& $S_z=0$ : $p(3d^{10})$  \\
($J=0$)       &  &  & $S_z=1$ : $\frac{2}{6}p(3d^8)$  &  &   \\
\hline
Two-orbital (ii) &  $S_z=0$ : $p(3d^6)$  %
		& $S_z=\frac{1}{2}$ : $p(3d^7)$ %
		& $S_z=1$ : $p(3d^8)$  %
		& $S_z=\frac{1}{2}$ : $p(3d^9)$ %
		& $S_z=0$ : $p(3d^{10})$  \\
($J$ large)       &  &  &   &  &   \\
\hline
Two-orbital (iii) &  $S_z=0$ : $p(3d^6)$  %
		& $S_z=\frac{1}{2}$ : $p(3d^7)$ %
		& $S_z=0$ : $\frac{3}{6}p(3d^8)$  %
		& $S_z=\frac{1}{2}$ : $p(3d^9)$ %
		& $S_z=0$ : $p(3d^{10})$  \\
($J$ intermediate)      &  &  & $S_z=1$ : $\frac{3}{6}p(3d^8)$  &  &   \\
\hline
\end{tabular}
\end{table*}

To clarify the type of system represented by these histograms it is useful to compare them to the spin statistics estimated from a simple one- and two-orbital model, taking into account the constraints from the occupation statistics measured in the full interacting systems.
We construct our two-orbital model by assuming that three of the Ni orbitals are fully occupied (the fluctuations to $3d^{5}$ and lower, not shown, are negligible). With this assumption the 0-4 electron sectors correspond to the Ni $3d^{6}$-$3d^{10}$ states in the full calculation, respectively. 
We consider the following three cases: (i) negligible Hund coupling ($J=0$) where the six possible 2-electron states contributing to $S_z=0$ and $S_z=1$ are equally probable, (ii) a large Hund coupling with only the high-spin configuration ($S_z$=1) allowed in the two-electron sector, and (iii) an intermediate coupling, where we assume the probabilities of the low- and high-spin states to be equal, $p(S_z=0)=p(S_z=1)$. Model (iii) slightly favors the high-spin state over model (i).
In the one-orbital model we additionally neglect the $3d^{6}$ and $3d^{7}$ states, and only represent the $3d^{8}$-$3d^{10}$ configurations by the 0-2 electron sectors. Since the configurations included in the one-orbital model still account for more than 90\% of the total weight measured in the full system, it should allow us to estimate if a single-orbital description is consistent with our results.

Figure~\ref{Fig:Spin model} plots the difference between the benchmark EDMFT results and the model estimates for the relative weights. The contributions to the spin $S_{z}=0,\,\frac{1}{2},\,1$ sectors of the various models are listed in Table~\ref{Table:Model spin}, and we use the calculated configuration probabilities shown in Fig.~\ref{Fig:Config Statistics} for the corresponding probabilities $p(3d^n)$.

The first observation we can make is that the $S_z=\frac{1}{2}$ weight is accurately captured by all models, although slightly better in the two-orbital picture than in the single-orbital one. The high-spin $S_z=1$ state, which is prominently populated in the EDMFT statistics, can obviously not be reproduced by the single-orbital model. However, more importantly, the single-orbital model significantly overestimates the $S_z=0$ weight (by almost a factor of 2) for both dopings. The two-orbital model with $J=0$ (model (i)), on the other hand, gives a relatively good agreement for both the $S_z=0$ (overestimated) and $S_z=1$ (underestimated) spin states. This is not the case for a large Hund coupling (model (ii)), which produces deviations on the same order as the single-orbital model, although in the opposite direction (underestimation of $S_z=0$).  

Taking into account competing effects which destabilize the high-spin state,  e. g. crystal-field level-splittings, we can surmise that the real situation is most adequately described by model (iii), which in the half-filled case assigns equal probabilities to the high-spin and low-spin configurations. Indeed, as shown in panels (c) and (g) of Fig.~\ref{Fig:Spin model}, this model provides the best agreement with the EDMFT results, both for the undoped and doped systems.
What these simple considerations show is that a single-orbital model cannot reproduce results which are consistent both with the occupation and spin statistics of the full model, while a two-orbital description is sufficient to reproduce both with good accuracy. 

We would also like to briefly comment on the pressure dependence of the predictions from the different models. Starting with the the one-orbital model, for the undoped system we see a slight improvement with increasing pressure, while the agreement for the doped system is equivalently worsened. This trend is shared by the prediction based on the two-orbital model without Hund coupling. The model with large coupling, although quantitatively not good, shows instead a comparable improvement for both the doped and undoped systems with increasing pressure. Model (iii), with intermediate Hund coupling effects, yields an improved description with increasing pressure for the doped system, while the undoped displays only very small changes without a clear trend.

Taken together, we interpret this result as further evidence for the multi-orbital nature of PrNiO$_2$; while a trend towards a single-band picture is discernible in the undoped system with increasing pressure, this is not the case for the doped system. Irrespective of the doping, a single-band description cannot capture the effect of Hund coupling, which strongly affects the population of the different local states. Its effect remains significant even if one focuses only on the fluctuations within the $3d_{x^2-y^2}$ orbital. 

\subsection{Energy levels}

 \begin{figure}[t]
  \includegraphics[width=0.49\textwidth]{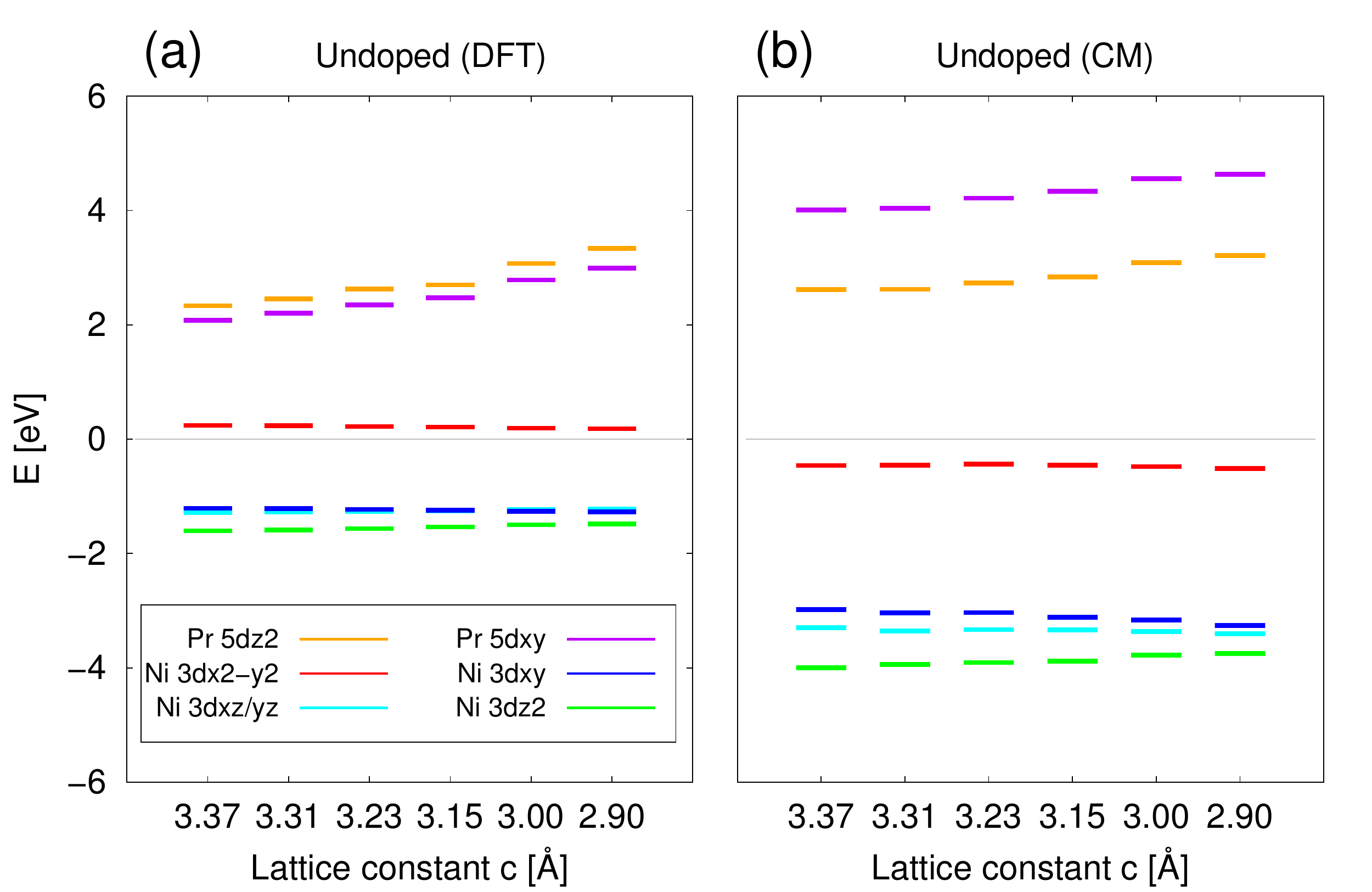}
   \includegraphics[width=0.49\textwidth]{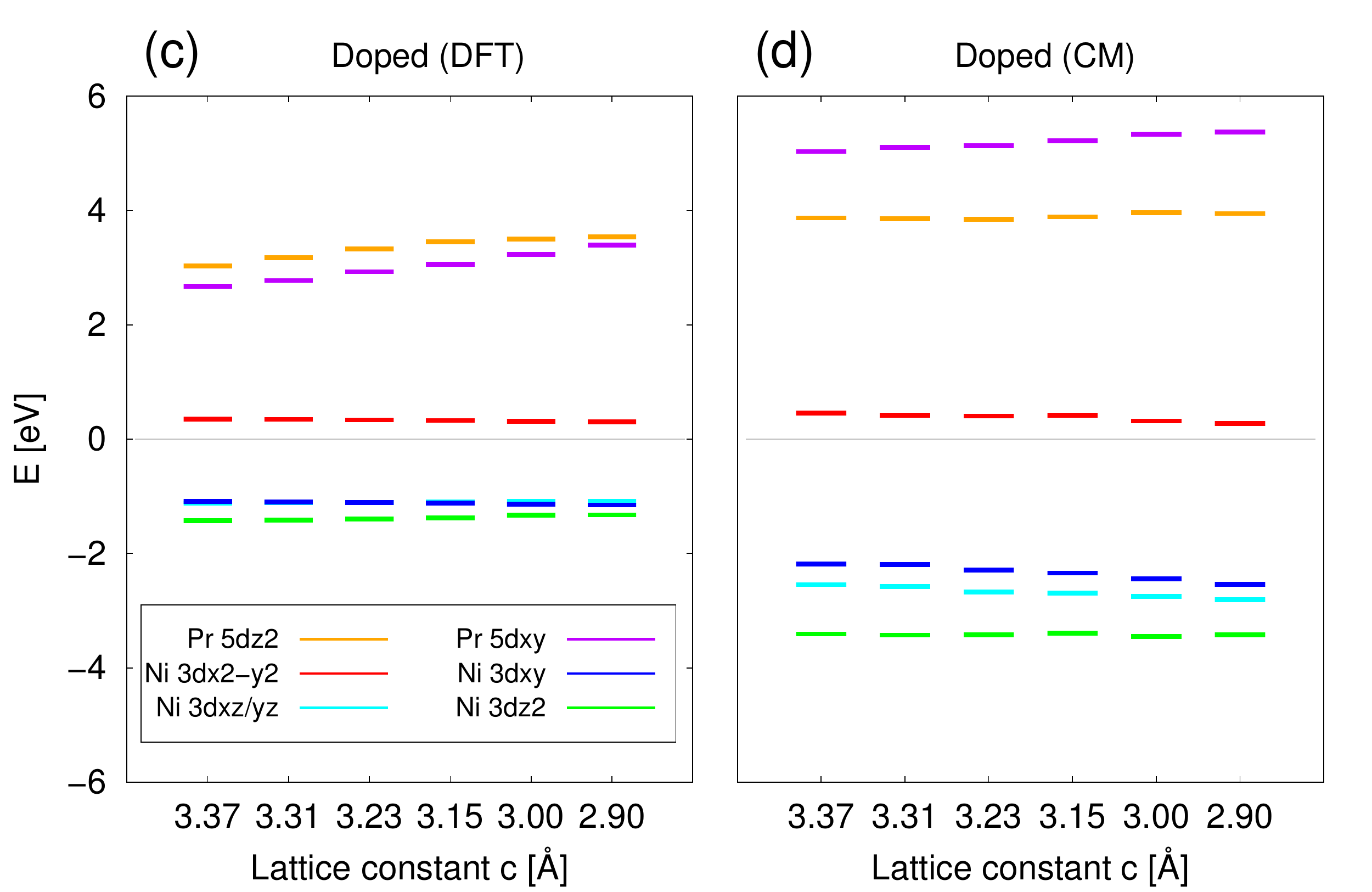}
  \caption{\label{Fig:Energy levels} Local energy level diagrams as a function of pressure (in terms of the lattice constant $c$) for the seven-orbital model. The different panels display results for the undoped compound derived from (a) DFT and (b) the center-of-mass of the spectral function in the interacting system, and for the doped system derived from (c) DFT and (d) the center-of-mass of the spectral function in the interacting system.}
 \end{figure}

We next discuss the renormalization of the local energy levels and the effect of pressure. In Fig.~\ref{Fig:Energy levels} we compare the DFT derived values, 
$\epsilon_\text{DFT}=\mathcal{H}_\text{DFT}({\bf R}=0)-\mu$, to the center of mass (CM) of the local spectral function $\epsilon_\textrm{CM}=\int\omega A(\omega)\mathrm{d}\omega$, where $A(\omega)$ is the local spectral function corresponding to the full interacting Green's function in Eq.~\eqref{Eq:Gint}.

Already on the DFT level we see a clear pressure effect on the local Pr energy levels, which are lifted up with increasing pressure for both dopings. The Ni orbitals instead remain approximately constant in energy, the only notable difference being a slight reordering of the $3d_{xz/yz}$ and $3d_{xy}$ energy levels towards high pressure. This contrasts with the CM derived local energy levels, which reveal that the almost degenerate Ni $3d_{xz/yz}$, $3d_{xy}$, and $3d_{z^2}$ DFT levels get split by correlation effects, while pressure acts to move them closer together.

We can furthermore relate the pressure dependence of the filling on the Pr site, which we discussed earlier, to the CM energy levels. For the undoped compound, we find the difference in energy to the Ni $3d_{x^2-y^2}$ level to increase with pressure, following the previously noted trend of a decrease in the occupation. This results in a reduced self-doping. Conversely, for the doped compound, in agreement with the initially lower occupation, the Pr are higher in energy, and they do not display the same relatively large shifts with pressure as we find without doping.

\subsection{Pressure effect on O $2p$} 
 
Up until this point we have focused only on the Ni and Pr manifolds, while omitting the O $p$ orbitals by integrating them out in the downfolding procedure following the DFT calculation.
In this section we also briefly discuss the pressure effects on the O $p$ local energy levels, derived by additionally including six oxygen-centered orbitals in the Wannierisation, yielding 13 orbitals in total. As expected, the larger energy window and number of bands reduces the spread for the Ni and Pr orbitals, and has the additional effect of lowering the local energies. We note, however, that the trends on the DFT level that we discussed previously remain unchanged, with the main difference being a slightly larger Ni 3d$_{xz/yz}$ -- Ni 3d$_{xy}$ splitting.

\begin{figure}[t]
 \includegraphics[width=0.45\textwidth]{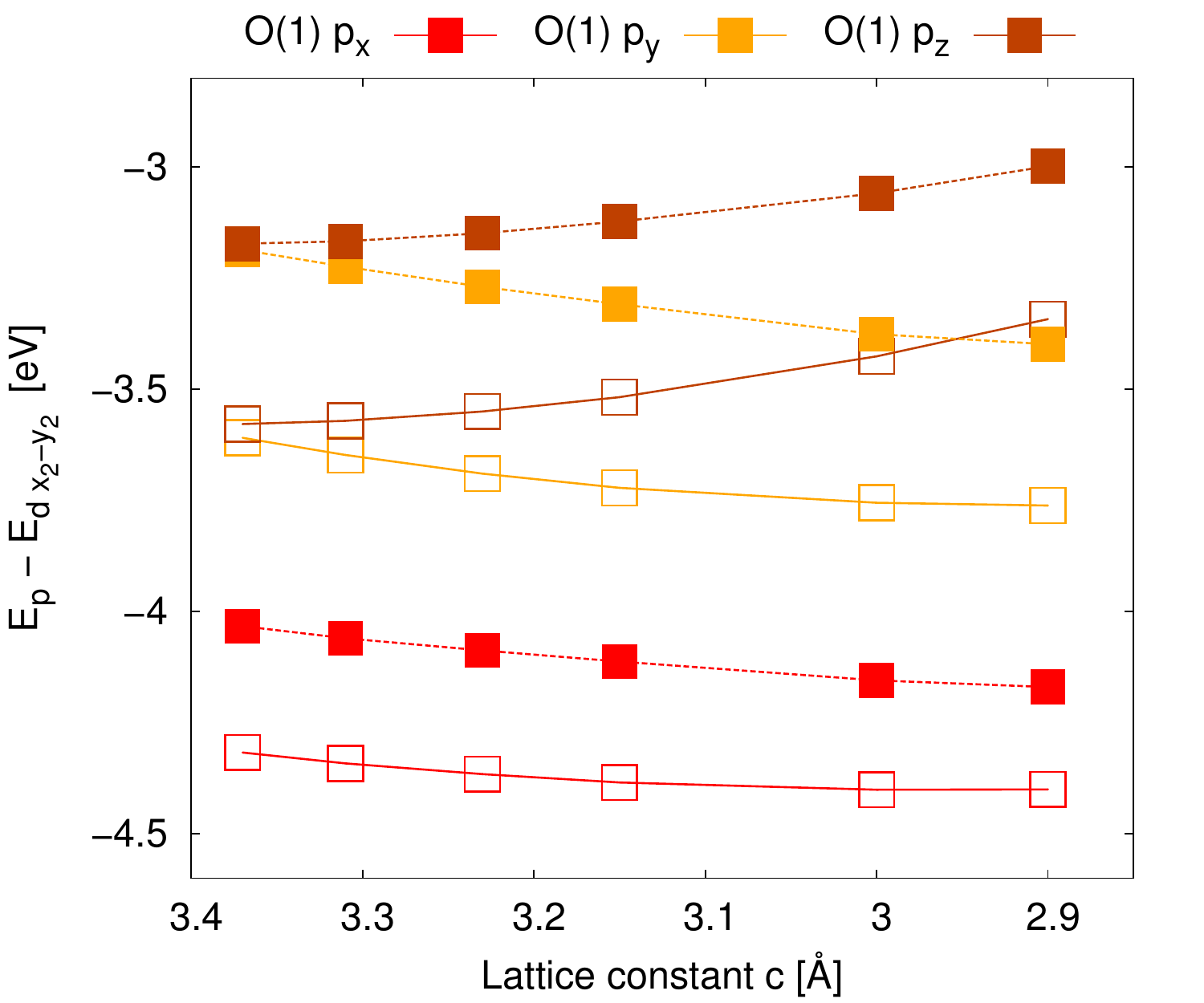}
  \caption{\label{Fig:Op energy levels} O(1) $p$ local  energies relative to Ni $3d_{x^2-y^2}$ as a function of pressure (represented in terms of the lattice constant $c$). Full lines with empty squares show the results for the undoped compound and the dashed lines with filled squares those for the doped compound. }
 \end{figure}

In Fig.~\ref{Fig:Op energy levels} we show the local energy levels $\epsilon_\text{DFT}$, relative to Ni $3d_{x^2-y^2}$ (which remains approximately pinned close to the Fermi energy), for the $2p$-like orbitals centered on the O(1) atom at (0.5,\,0,\,0). The equivalent Wannier functions for O(2) at (0,\,0.5,\,0) are related as: O(1)~$p_x$ -- O(2)~$p_y$ and O(1)~$p_y$ -- O(2)~$p_x$, and the out-of-plane $p_z$ for the two oxygens are equivalent. 

Overall we note that the O~$2p$ orbitals in the doped compound are higher in energy, compared to the undoped ones, on the order of $\sim$0.5 eV.
Additionally, we observe again two different systematic changes with increasing pressure; the in-plane orbitals ($p_x$ and $p_y$) for both the doped and undoped systems are shifted down with pressure, while interestingly, the opposite trend is observed for the $p_z$ orbitals which are shifted up in energy instead.
As the in-plane orbitals are expected to hybridize more strongly with the Ni $3d_{x^2-y^2}$ orbital, this suggests a reduced involvement of the oxygen orbitals (with increasing pressure) in the mechanism underpinning the reported increase in $T_c$, at least on the DFT level. Future work taking into account also the O $2p$ orbitals in the self-consistency would be needed to settle this question.

\section{\label{sec:Summary}Summary and Conclusions}

To gain insights into the effect of physical pressure on the electronic structure of infinite layered nickelates, we used the parameter-free $GW$+EDMFT method to calculate orbital occupations, effective energy levels and effective interaction parameters. 
Our investigation focused on PrNiO$_2$, for which a substantial increase of $T_c$ with increasing pressure has recently been reported in experiments.\cite{Wang2022} The $GW$+EDMFT results also provided insights into the question of the single- versus multi-orbital nature of the system, and how this picture is affected by the application of pressure. The numerical calculations revealed quantitatively small changes, but clear systematic trends. One of the most interesting findings is the subtle interplay between chemical doping and pressure, which results in qualitatively different pressure evolutions for different doping levels, and different physical pictures. 

We found several indications that the undoped system becomes more single-band-like at higher pressures. In the orbital occupations, one observes charge transfer mostly from Pr to the filled Ni orbitals, leaving the Ni $3d_{x^2-y^2}$ orbital essentially pinned at a fixed filling.
Also the results for the occupation and spin statistics in the high pressure region become increasingly more consistent with the behavior expected for a single-band system.
The doped compound, however, does not display the changes that one would naively associate with an increasingly single-band-like picture. The charge transfer resulting from the application of pressure to a large extent occurs within the Ni $3d$ manifold. In particular we found a non-monotonic change in the $3d_{z^2}$ occupation, which is responsible for moving the $3d_{x^2-y^2}$ orbital towards half-filling in the high-pressure regime. Such effects cannot be captured within a single-orbital picture. We also showed that at least a two-orbital model is needed to capture the dominant fluctuations in the occupation and spin states.
The reason is the significant weight of high-spin states, which affects the fluctuations even in the undoped system, where occupations compatible with an increasingly single-band-like picture under pressure are found.
Our analysis showed that a two-orbital model, with competing effects between Hund coupling and, e.g., crystal-field splitting, is the minimal model which satisfactorily reproduces the relevant sectors of the EDMFT spin statistics of the full 7-orbital low-energy model.

The calculation of the effective local energy levels of the orbitals revealed that the increase in the level splittings between Ni $3d_{x^2-y^2}$ and the Pr orbitals is consistent with a reduced self-doping from Pr under pressure in the undoped system, while the changes are less noteworthy in the doped system. This is in agreement with our analysis of the orbital occupations. Within DFT, these relative shifts appear to be relatively large also in the doped system and could therefore overestimate the pressure effects.
Our DFT analysis of the influence of pressure on the O~$2p$ orbitals indicates that they are most likely not fundamental for understanding the pressure-induced increase in $T_c$, as the in-plane orbitals are systematically lowered in energy relative to Ni $3d_{x^2-y^2}$ orbitals which they would be expected to hybridize with.
Including the O orbitals in the low-energy space to self-consistently capture their effect and fully clarify their role would be an interesting task, which is left for future work.

The interaction parameters in the low-energy model exhibit small, but systematic changes over the considered pressure range.
Especially the calculated intra-orbital interaction is of interest due to its possible relevance for the $T_c$ estimate in a single-band picture based on $D\Gamma A$.\cite{Kitatani2020} While the cRPA estimate and the self-consistently computed $\mathcal{U}$ show similar trends,
the latter value is reduced as a result of low-energy screening processes, which again highlights the importance of an accurate treatment of the low-energy physics. 
Our calculations revealed that the expected decrease in the $U/t$ ratio under pressure is counteracted by a substantial increase in the bare interaction with pressure. Depending on how we estimate the hopping parameter $t$, either based on the Wannier Hamiltonian or the bandwidth, we can reach different conclusions. 
The estimate based on the Wannier Hamiltonian suggests a clear increase of $U/t$ under pressure.
In the undoped system, the estimate based on the bandwidth does lead to a decrease in $U/t$, while intriguingly this is again not the case for the doped system, which yields an approximately constant or even slightly increasing $U/t$. 
These results suggest that an explanation of the experimentally observed $T_c$ trend based on the {\it a priori} expected evolution of $U/t$ can be misleading.
However, further investigations of the screening and band widening effects are needed to settle the question.

In the cuprate context, single-band behavior (in the sense of a weak hybridization between the $d_{x^2-y^2}$ and $d_{z^2}$ orbitals) has been linked to high $T_c$ values.\cite{Sakakibara2010} For this reason, it is tempting to interpret the increased $T_c$ in Pr$_{1-x}$Sr$_x$NiO$_2$ under pressure as a consequence of the pressure-dependent level diagram (Fig.~\ref{Fig:Energy levels}), which indicates an increasing decoupling of the $d_{x^2-y^2}$ orbital from the other orbitals, both at the DFT and $GW$+EDMFT level.
While this trend looks consistent, the quantitative shifts in the doped interacting system are however small and may not by themselves explain the large enhancement of $T_c$ seen in the experiments. 
Further experimental and theoretical investigations of the pressure effects at different doping levels, including the undoped compound, in PrNiO$_2$ and other nickelate superconductors would be helpful to establish systematic trends and to clarify the interplay between doping and physical pressure. 
Separate investigations of the pairing mechanism, which take into account the doping and pressure-dependent modifications of the screening and the level shifts revealed in this study, as well as the multi-orbital nature of the material, are needed to clarify the effects of pressure on superconductivity.

\begin{acknowledgments}
The calculations have been performed on the Beo05 cluster at the University of Fribourg. 
This work was supported by ERC Consolidator Grant No. 724103 and by the Swiss National Science Foundation via Grant No. 200021-196966.
\end{acknowledgments}

\bibliography{paper.bib}

\begin{thebibliography}{58}%
\makeatletter
\providecommand \@ifxundefined [1]{%
 \@ifx{#1\undefined}
}%
\providecommand \@ifnum [1]{%
 \ifnum #1\expandafter \@firstoftwo
 \else \expandafter \@secondoftwo
 \fi
}%
\providecommand \@ifx [1]{%
 \ifx #1\expandafter \@firstoftwo
 \else \expandafter \@secondoftwo
 \fi
}%
\providecommand \natexlab [1]{#1}%
\providecommand \enquote  [1]{``#1''}%
\providecommand \bibnamefont  [1]{#1}%
\providecommand \bibfnamefont [1]{#1}%
\providecommand \citenamefont [1]{#1}%
\providecommand \href@noop [0]{\@secondoftwo}%
\providecommand \href [0]{\begingroup \@sanitize@url \@href}%
\providecommand \@href[1]{\@@startlink{#1}\@@href}%
\providecommand \@@href[1]{\endgroup#1\@@endlink}%
\providecommand \@sanitize@url [0]{\catcode `\\12\catcode `\$12\catcode
  `\&12\catcode `\#12\catcode `\^12\catcode `\_12\catcode `\%12\relax}%
\providecommand \@@startlink[1]{}%
\providecommand \@@endlink[0]{}%
\providecommand \url  [0]{\begingroup\@sanitize@url \@url }%
\providecommand \@url [1]{\endgroup\@href {#1}{\urlprefix }}%
\providecommand \urlprefix  [0]{URL }%
\providecommand \Eprint [0]{\href }%
\providecommand \doibase [0]{http://dx.doi.org/}%
\providecommand \selectlanguage [0]{\@gobble}%
\providecommand \bibinfo  [0]{\@secondoftwo}%
\providecommand \bibfield  [0]{\@secondoftwo}%
\providecommand \translation [1]{[#1]}%
\providecommand \BibitemOpen [0]{}%
\providecommand \bibitemStop [0]{}%
\providecommand \bibitemNoStop [0]{.\EOS\space}%
\providecommand \EOS [0]{\spacefactor3000\relax}%
\providecommand \BibitemShut  [1]{\csname bibitem#1\endcsname}%
\let\auto@bib@innerbib\@empty
\bibitem [{\citenamefont {Li}\ \emph {et~al.}(2019)\citenamefont {Li},
  \citenamefont {Lee}, \citenamefont {Wang}, \citenamefont {Osada},
  \citenamefont {Crossley}, \citenamefont {Lee}, \citenamefont {Cui},
  \citenamefont {Hikita},\ and\ \citenamefont {Hwang}}]{Li2019}%
  \BibitemOpen
  \bibfield  {author} {\bibinfo {author} {\bibfnamefont {D.}~\bibnamefont
  {Li}}, \bibinfo {author} {\bibfnamefont {K.}~\bibnamefont {Lee}}, \bibinfo
  {author} {\bibfnamefont {B.~Y.}\ \bibnamefont {Wang}}, \bibinfo {author}
  {\bibfnamefont {M.}~\bibnamefont {Osada}}, \bibinfo {author} {\bibfnamefont
  {S.}~\bibnamefont {Crossley}}, \bibinfo {author} {\bibfnamefont {H.~R.}\
  \bibnamefont {Lee}}, \bibinfo {author} {\bibfnamefont {Y.}~\bibnamefont
  {Cui}}, \bibinfo {author} {\bibfnamefont {Y.}~\bibnamefont {Hikita}}, \ and\
  \bibinfo {author} {\bibfnamefont {H.~Y.}\ \bibnamefont {Hwang}},\ }\href
  {\doibase 10.1038/s41586-019-1496-5} {\bibfield  {journal} {\bibinfo
  {journal} {Nature}\ }\textbf {\bibinfo {volume} {572}},\ \bibinfo {pages}
  {624} (\bibinfo {year} {2019})}\BibitemShut {NoStop}%
\bibitem [{\citenamefont {Osada}\ \emph
  {et~al.}(2020{\natexlab{a}})\citenamefont {Osada}, \citenamefont {Wang},
  \citenamefont {Goodge}, \citenamefont {Lee}, \citenamefont {Yoon},
  \citenamefont {Sakuma}, \citenamefont {Li}, \citenamefont {Miura},
  \citenamefont {Kourkoutis},\ and\ \citenamefont {Hwang}}]{Osada2020a}%
  \BibitemOpen
  \bibfield  {author} {\bibinfo {author} {\bibfnamefont {M.}~\bibnamefont
  {Osada}}, \bibinfo {author} {\bibfnamefont {B.~Y.}\ \bibnamefont {Wang}},
  \bibinfo {author} {\bibfnamefont {B.~H.}\ \bibnamefont {Goodge}}, \bibinfo
  {author} {\bibfnamefont {K.}~\bibnamefont {Lee}}, \bibinfo {author}
  {\bibfnamefont {H.}~\bibnamefont {Yoon}}, \bibinfo {author} {\bibfnamefont
  {K.}~\bibnamefont {Sakuma}}, \bibinfo {author} {\bibfnamefont
  {D.}~\bibnamefont {Li}}, \bibinfo {author} {\bibfnamefont {M.}~\bibnamefont
  {Miura}}, \bibinfo {author} {\bibfnamefont {L.~F.}\ \bibnamefont
  {Kourkoutis}}, \ and\ \bibinfo {author} {\bibfnamefont {H.~Y.}\ \bibnamefont
  {Hwang}},\ }\href {\doibase 10.1021/acs.nanolett.0c01392} {\bibfield
  {journal} {\bibinfo  {journal} {Nano Letters}\ }\textbf {\bibinfo {volume}
  {20}},\ \bibinfo {pages} {5735} (\bibinfo {year}
  {2020}{\natexlab{a}})}\BibitemShut {NoStop}%
\bibitem [{\citenamefont {Osada}\ \emph
  {et~al.}(2020{\natexlab{b}})\citenamefont {Osada}, \citenamefont {Wang},
  \citenamefont {Lee}, \citenamefont {Li},\ and\ \citenamefont
  {Hwang}}]{Osada2020b}%
  \BibitemOpen
  \bibfield  {author} {\bibinfo {author} {\bibfnamefont {M.}~\bibnamefont
  {Osada}}, \bibinfo {author} {\bibfnamefont {B.~Y.}\ \bibnamefont {Wang}},
  \bibinfo {author} {\bibfnamefont {K.}~\bibnamefont {Lee}}, \bibinfo {author}
  {\bibfnamefont {D.}~\bibnamefont {Li}}, \ and\ \bibinfo {author}
  {\bibfnamefont {H.~Y.}\ \bibnamefont {Hwang}},\ }\href {\doibase
  10.1103/PhysRevMaterials.4.121801} {\bibfield  {journal} {\bibinfo  {journal}
  {Phys. Rev. Materials}\ }\textbf {\bibinfo {volume} {4}},\ \bibinfo {pages}
  {121801} (\bibinfo {year} {2020}{\natexlab{b}})}\BibitemShut {NoStop}%
\bibitem [{\citenamefont {Osada}\ \emph {et~al.}(2021)\citenamefont {Osada},
  \citenamefont {Wang}, \citenamefont {Goodge}, \citenamefont {Harvey},
  \citenamefont {Lee}, \citenamefont {Li}, \citenamefont {Kourkoutis},\ and\
  \citenamefont {Hwang}}]{Osada2021}%
  \BibitemOpen
  \bibfield  {author} {\bibinfo {author} {\bibfnamefont {M.}~\bibnamefont
  {Osada}}, \bibinfo {author} {\bibfnamefont {B.~Y.}\ \bibnamefont {Wang}},
  \bibinfo {author} {\bibfnamefont {B.~H.}\ \bibnamefont {Goodge}}, \bibinfo
  {author} {\bibfnamefont {S.~P.}\ \bibnamefont {Harvey}}, \bibinfo {author}
  {\bibfnamefont {K.}~\bibnamefont {Lee}}, \bibinfo {author} {\bibfnamefont
  {D.}~\bibnamefont {Li}}, \bibinfo {author} {\bibfnamefont {L.~F.}\
  \bibnamefont {Kourkoutis}}, \ and\ \bibinfo {author} {\bibfnamefont {H.~Y.}\
  \bibnamefont {Hwang}},\ }\href {\doibase
  https://doi.org/10.1002/adma.202104083} {\bibfield  {journal} {\bibinfo
  {journal} {Advanced Materials}\ }\textbf {\bibinfo {volume} {33}},\ \bibinfo
  {pages} {2104083} (\bibinfo {year} {2021})}\BibitemShut {NoStop}%
\bibitem [{\citenamefont {Zeng}\ \emph {et~al.}(2022)\citenamefont {Zeng},
  \citenamefont {Li}, \citenamefont {Chow}, \citenamefont {Cao}, \citenamefont
  {Zhang}, \citenamefont {Tang}, \citenamefont {Yin}, \citenamefont {Lim},
  \citenamefont {Hu}, \citenamefont {Yang},\ and\ \citenamefont
  {Ariando}}]{Zeng2022}%
  \BibitemOpen
  \bibfield  {author} {\bibinfo {author} {\bibfnamefont {S.}~\bibnamefont
  {Zeng}}, \bibinfo {author} {\bibfnamefont {C.}~\bibnamefont {Li}}, \bibinfo
  {author} {\bibfnamefont {L.~E.}\ \bibnamefont {Chow}}, \bibinfo {author}
  {\bibfnamefont {Y.}~\bibnamefont {Cao}}, \bibinfo {author} {\bibfnamefont
  {Z.}~\bibnamefont {Zhang}}, \bibinfo {author} {\bibfnamefont {C.~S.}\
  \bibnamefont {Tang}}, \bibinfo {author} {\bibfnamefont {X.}~\bibnamefont
  {Yin}}, \bibinfo {author} {\bibfnamefont {Z.~S.}\ \bibnamefont {Lim}},
  \bibinfo {author} {\bibfnamefont {J.}~\bibnamefont {Hu}}, \bibinfo {author}
  {\bibfnamefont {P.}~\bibnamefont {Yang}}, \ and\ \bibinfo {author}
  {\bibfnamefont {A.}~\bibnamefont {Ariando}},\ }\href {\doibase
  10.1126/sciadv.abl9927} {\bibfield  {journal} {\bibinfo  {journal} {Science
  Advances}\ }\textbf {\bibinfo {volume} {8}} (\bibinfo {year} {2022}),\
  10.1126/sciadv.abl9927}\BibitemShut {NoStop}%
\bibitem [{\citenamefont {Pan}\ \emph {et~al.}(2022)\citenamefont {Pan},
  \citenamefont {Ferenc~Segedin}, \citenamefont {LaBollita}, \citenamefont
  {Song}, \citenamefont {Nica}, \citenamefont {Goodge}, \citenamefont {Pierce},
  \citenamefont {Doyle}, \citenamefont {Novakov}, \citenamefont
  {C{\'o}rdova~Carrizales}, \citenamefont {N'Diaye}, \citenamefont {Shafer},
  \citenamefont {Paik}, \citenamefont {Heron}, \citenamefont {Mason},
  \citenamefont {Yacoby}, \citenamefont {Kourkoutis}, \citenamefont {Erten},
  \citenamefont {Brooks}, \citenamefont {Botana},\ and\ \citenamefont
  {Mundy}}]{Pan2021}%
  \BibitemOpen
  \bibfield  {author} {\bibinfo {author} {\bibfnamefont {G.~A.}\ \bibnamefont
  {Pan}}, \bibinfo {author} {\bibfnamefont {D.}~\bibnamefont {Ferenc~Segedin}},
  \bibinfo {author} {\bibfnamefont {H.}~\bibnamefont {LaBollita}}, \bibinfo
  {author} {\bibfnamefont {Q.}~\bibnamefont {Song}}, \bibinfo {author}
  {\bibfnamefont {E.~M.}\ \bibnamefont {Nica}}, \bibinfo {author}
  {\bibfnamefont {B.~H.}\ \bibnamefont {Goodge}}, \bibinfo {author}
  {\bibfnamefont {A.~T.}\ \bibnamefont {Pierce}}, \bibinfo {author}
  {\bibfnamefont {S.}~\bibnamefont {Doyle}}, \bibinfo {author} {\bibfnamefont
  {S.}~\bibnamefont {Novakov}}, \bibinfo {author} {\bibfnamefont
  {D.}~\bibnamefont {C{\'o}rdova~Carrizales}}, \bibinfo {author} {\bibfnamefont
  {A.~T.}\ \bibnamefont {N'Diaye}}, \bibinfo {author} {\bibfnamefont
  {P.}~\bibnamefont {Shafer}}, \bibinfo {author} {\bibfnamefont
  {H.}~\bibnamefont {Paik}}, \bibinfo {author} {\bibfnamefont {J.~T.}\
  \bibnamefont {Heron}}, \bibinfo {author} {\bibfnamefont {J.~A.}\ \bibnamefont
  {Mason}}, \bibinfo {author} {\bibfnamefont {A.}~\bibnamefont {Yacoby}},
  \bibinfo {author} {\bibfnamefont {L.~F.}\ \bibnamefont {Kourkoutis}},
  \bibinfo {author} {\bibfnamefont {O.}~\bibnamefont {Erten}}, \bibinfo
  {author} {\bibfnamefont {C.~M.}\ \bibnamefont {Brooks}}, \bibinfo {author}
  {\bibfnamefont {A.~S.}\ \bibnamefont {Botana}}, \ and\ \bibinfo {author}
  {\bibfnamefont {J.~A.}\ \bibnamefont {Mundy}},\ }\href {\doibase
  10.1038/s41563-021-01142-9} {\bibfield  {journal} {\bibinfo  {journal}
  {Nature Materials}\ }\textbf {\bibinfo {volume} {21}},\ \bibinfo {pages}
  {160} (\bibinfo {year} {2022})}\BibitemShut {NoStop}%
\bibitem [{\citenamefont {Nomura}\ \emph {et~al.}(2019)\citenamefont {Nomura},
  \citenamefont {Hirayama}, \citenamefont {Tadano}, \citenamefont {Yoshimoto},
  \citenamefont {Nakamura},\ and\ \citenamefont {Arita}}]{Nomura2019}%
  \BibitemOpen
  \bibfield  {author} {\bibinfo {author} {\bibfnamefont {Y.}~\bibnamefont
  {Nomura}}, \bibinfo {author} {\bibfnamefont {M.}~\bibnamefont {Hirayama}},
  \bibinfo {author} {\bibfnamefont {T.}~\bibnamefont {Tadano}}, \bibinfo
  {author} {\bibfnamefont {Y.}~\bibnamefont {Yoshimoto}}, \bibinfo {author}
  {\bibfnamefont {K.}~\bibnamefont {Nakamura}}, \ and\ \bibinfo {author}
  {\bibfnamefont {R.}~\bibnamefont {Arita}},\ }\href {\doibase
  10.1103/PhysRevB.100.205138} {\bibfield  {journal} {\bibinfo  {journal}
  {Phys. Rev. B}\ }\textbf {\bibinfo {volume} {100}},\ \bibinfo {pages}
  {205138} (\bibinfo {year} {2019})}\BibitemShut {NoStop}%
\bibitem [{\citenamefont {Kitatani}\ \emph {et~al.}(2020)\citenamefont
  {Kitatani}, \citenamefont {Si}, \citenamefont {Janson}, \citenamefont
  {Arita}, \citenamefont {Zhong},\ and\ \citenamefont {Held}}]{Kitatani2020}%
  \BibitemOpen
  \bibfield  {author} {\bibinfo {author} {\bibfnamefont {M.}~\bibnamefont
  {Kitatani}}, \bibinfo {author} {\bibfnamefont {L.}~\bibnamefont {Si}},
  \bibinfo {author} {\bibfnamefont {O.}~\bibnamefont {Janson}}, \bibinfo
  {author} {\bibfnamefont {R.}~\bibnamefont {Arita}}, \bibinfo {author}
  {\bibfnamefont {Z.}~\bibnamefont {Zhong}}, \ and\ \bibinfo {author}
  {\bibfnamefont {K.}~\bibnamefont {Held}},\ }\href {\doibase
  10.1038/s41535-020-00260-y} {\bibfield  {journal} {\bibinfo  {journal} {npj
  Quantum Mater.}\ }\textbf {\bibinfo {volume} {5}},\ \bibinfo {pages} {59}
  (\bibinfo {year} {2020})}\BibitemShut {NoStop}%
\bibitem [{\citenamefont {Karp}\ \emph
  {et~al.}(2020{\natexlab{a}})\citenamefont {Karp}, \citenamefont {Botana},
  \citenamefont {Norman}, \citenamefont {Park}, \citenamefont {Zingl},\ and\
  \citenamefont {Millis}}]{Karp2020a}%
  \BibitemOpen
  \bibfield  {author} {\bibinfo {author} {\bibfnamefont {J.}~\bibnamefont
  {Karp}}, \bibinfo {author} {\bibfnamefont {A.~S.}\ \bibnamefont {Botana}},
  \bibinfo {author} {\bibfnamefont {M.~R.}\ \bibnamefont {Norman}}, \bibinfo
  {author} {\bibfnamefont {H.}~\bibnamefont {Park}}, \bibinfo {author}
  {\bibfnamefont {M.}~\bibnamefont {Zingl}}, \ and\ \bibinfo {author}
  {\bibfnamefont {A.}~\bibnamefont {Millis}},\ }\href {\doibase
  10.1103/PhysRevX.10.021061} {\bibfield  {journal} {\bibinfo  {journal} {Phys.
  Rev. X}\ }\textbf {\bibinfo {volume} {10}},\ \bibinfo {pages} {021061}
  (\bibinfo {year} {2020}{\natexlab{a}})}\BibitemShut {NoStop}%
\bibitem [{\citenamefont {Karp}\ \emph
  {et~al.}(2020{\natexlab{b}})\citenamefont {Karp}, \citenamefont {Hampel},
  \citenamefont {Zingl}, \citenamefont {Botana}, \citenamefont {Park},
  \citenamefont {Norman},\ and\ \citenamefont {Millis}}]{Karp2020b}%
  \BibitemOpen
  \bibfield  {author} {\bibinfo {author} {\bibfnamefont {J.}~\bibnamefont
  {Karp}}, \bibinfo {author} {\bibfnamefont {A.}~\bibnamefont {Hampel}},
  \bibinfo {author} {\bibfnamefont {M.}~\bibnamefont {Zingl}}, \bibinfo
  {author} {\bibfnamefont {A.~S.}\ \bibnamefont {Botana}}, \bibinfo {author}
  {\bibfnamefont {H.}~\bibnamefont {Park}}, \bibinfo {author} {\bibfnamefont
  {M.~R.}\ \bibnamefont {Norman}}, \ and\ \bibinfo {author} {\bibfnamefont
  {A.~J.}\ \bibnamefont {Millis}},\ }\href {\doibase
  10.1103/PhysRevB.102.245130} {\bibfield  {journal} {\bibinfo  {journal}
  {Phys. Rev. B}\ }\textbf {\bibinfo {volume} {102}},\ \bibinfo {pages}
  {245130} (\bibinfo {year} {2020}{\natexlab{b}})}\BibitemShut {NoStop}%
\bibitem [{\citenamefont {Higashi}\ \emph {et~al.}(2021)\citenamefont
  {Higashi}, \citenamefont {Winder}, \citenamefont {Kune\ifmmode~\check{s}\else
  \v{s}\fi{}},\ and\ \citenamefont {Hariki}}]{Higashi2021}%
  \BibitemOpen
  \bibfield  {author} {\bibinfo {author} {\bibfnamefont {K.}~\bibnamefont
  {Higashi}}, \bibinfo {author} {\bibfnamefont {M.}~\bibnamefont {Winder}},
  \bibinfo {author} {\bibfnamefont {J.}~\bibnamefont
  {Kune\ifmmode~\check{s}\else \v{s}\fi{}}}, \ and\ \bibinfo {author}
  {\bibfnamefont {A.}~\bibnamefont {Hariki}},\ }\href {\doibase
  10.1103/PhysRevX.11.041009} {\bibfield  {journal} {\bibinfo  {journal} {Phys.
  Rev. X}\ }\textbf {\bibinfo {volume} {11}},\ \bibinfo {pages} {041009}
  (\bibinfo {year} {2021})}\BibitemShut {NoStop}%
\bibitem [{\citenamefont {Lechermann}(2020{\natexlab{a}})}]{Lechermann2020a}%
  \BibitemOpen
  \bibfield  {author} {\bibinfo {author} {\bibfnamefont {F.}~\bibnamefont
  {Lechermann}},\ }\href {\doibase 10.1103/PhysRevB.101.081110} {\bibfield
  {journal} {\bibinfo  {journal} {Phys. Rev. B}\ }\textbf {\bibinfo {volume}
  {101}},\ \bibinfo {pages} {081110} (\bibinfo {year}
  {2020}{\natexlab{a}})}\BibitemShut {NoStop}%
\bibitem [{\citenamefont {Werner}\ and\ \citenamefont
  {Hoshino}(2020)}]{Werner2020}%
  \BibitemOpen
  \bibfield  {author} {\bibinfo {author} {\bibfnamefont {P.}~\bibnamefont
  {Werner}}\ and\ \bibinfo {author} {\bibfnamefont {S.}~\bibnamefont
  {Hoshino}},\ }\href {\doibase 10.1103/PhysRevB.101.041104} {\bibfield
  {journal} {\bibinfo  {journal} {Phys. Rev. B}\ }\textbf {\bibinfo {volume}
  {101}},\ \bibinfo {pages} {041104(R)} (\bibinfo {year} {2020})}\BibitemShut
  {NoStop}%
\bibitem [{\citenamefont {Lechermann}(2020{\natexlab{b}})}]{Lechermann2020b}%
  \BibitemOpen
  \bibfield  {author} {\bibinfo {author} {\bibfnamefont {F.}~\bibnamefont
  {Lechermann}},\ }\href {\doibase 10.1103/PhysRevX.10.041002} {\bibfield
  {journal} {\bibinfo  {journal} {Phys. Rev. X}\ }\textbf {\bibinfo {volume}
  {10}},\ \bibinfo {pages} {041002} (\bibinfo {year}
  {2020}{\natexlab{b}})}\BibitemShut {NoStop}%
\bibitem [{\citenamefont {Petocchi}\ \emph
  {et~al.}(2020{\natexlab{a}})\citenamefont {Petocchi}, \citenamefont
  {Christiansson}, \citenamefont {Nilsson}, \citenamefont {Aryasetiawan},\ and\
  \citenamefont {Werner}}]{Petocchi2020b}%
  \BibitemOpen
  \bibfield  {author} {\bibinfo {author} {\bibfnamefont {F.}~\bibnamefont
  {Petocchi}}, \bibinfo {author} {\bibfnamefont {V.}~\bibnamefont
  {Christiansson}}, \bibinfo {author} {\bibfnamefont {F.}~\bibnamefont
  {Nilsson}}, \bibinfo {author} {\bibfnamefont {F.}~\bibnamefont
  {Aryasetiawan}}, \ and\ \bibinfo {author} {\bibfnamefont {P.}~\bibnamefont
  {Werner}},\ }\href {\doibase 10.1103/PhysRevX.10.041047} {\bibfield
  {journal} {\bibinfo  {journal} {Phys. Rev. X}\ }\textbf {\bibinfo {volume}
  {10}},\ \bibinfo {pages} {041047} (\bibinfo {year}
  {2020}{\natexlab{a}})}\BibitemShut {NoStop}%
\bibitem [{\citenamefont {Kang}\ \emph {et~al.}(2020)\citenamefont {Kang},
  \citenamefont {Melnick}, \citenamefont {Semon}, \citenamefont {Ryee},
  \citenamefont {Han}, \citenamefont {Kotliar},\ and\ \citenamefont
  {Choi}}]{Kang2020}%
  \BibitemOpen
  \bibfield  {author} {\bibinfo {author} {\bibfnamefont {B.}~\bibnamefont
  {Kang}}, \bibinfo {author} {\bibfnamefont {C.}~\bibnamefont {Melnick}},
  \bibinfo {author} {\bibfnamefont {P.}~\bibnamefont {Semon}}, \bibinfo
  {author} {\bibfnamefont {S.}~\bibnamefont {Ryee}}, \bibinfo {author}
  {\bibfnamefont {M.~J.}\ \bibnamefont {Han}}, \bibinfo {author} {\bibfnamefont
  {G.}~\bibnamefont {Kotliar}}, \ and\ \bibinfo {author} {\bibfnamefont
  {S.}~\bibnamefont {Choi}},\ }\href {\doibase 10.48550/ARXIV.2007.14610}
  {\enquote {\bibinfo {title} {Infinite-layer nickelates as ni-eg hund's
  metals},}\ } (\bibinfo {year} {2020}),\ \bibinfo {note}
  {arXiv:2007.14610}\BibitemShut {NoStop}%
\bibitem [{\citenamefont {Wang}\ \emph {et~al.}(2020)\citenamefont {Wang},
  \citenamefont {Kang}, \citenamefont {Miao},\ and\ \citenamefont
  {Kotliar}}]{Wang2020}%
  \BibitemOpen
  \bibfield  {author} {\bibinfo {author} {\bibfnamefont {Y.}~\bibnamefont
  {Wang}}, \bibinfo {author} {\bibfnamefont {C.-J.}\ \bibnamefont {Kang}},
  \bibinfo {author} {\bibfnamefont {H.}~\bibnamefont {Miao}}, \ and\ \bibinfo
  {author} {\bibfnamefont {G.}~\bibnamefont {Kotliar}},\ }\href {\doibase
  10.1103/PhysRevB.102.161118} {\bibfield  {journal} {\bibinfo  {journal}
  {Phys. Rev. B}\ }\textbf {\bibinfo {volume} {102}},\ \bibinfo {pages}
  {161118} (\bibinfo {year} {2020})}\BibitemShut {NoStop}%
\bibitem [{\citenamefont {Lee}\ and\ \citenamefont {Pickett}(2004)}]{Lee2004}%
  \BibitemOpen
  \bibfield  {author} {\bibinfo {author} {\bibfnamefont {K.-W.}\ \bibnamefont
  {Lee}}\ and\ \bibinfo {author} {\bibfnamefont {W.~E.}\ \bibnamefont
  {Pickett}},\ }\href {\doibase 10.1103/PhysRevB.70.165109} {\bibfield
  {journal} {\bibinfo  {journal} {Phys. Rev. B}\ }\textbf {\bibinfo {volume}
  {70}},\ \bibinfo {pages} {165109} (\bibinfo {year} {2004})}\BibitemShut
  {NoStop}%
\bibitem [{\citenamefont {Gu}\ \emph {et~al.}(2020)\citenamefont {Gu},
  \citenamefont {Zhu}, \citenamefont {Wang}, \citenamefont {Hu},\ and\
  \citenamefont {Chen}}]{Gu2020}%
  \BibitemOpen
  \bibfield  {author} {\bibinfo {author} {\bibfnamefont {Y.}~\bibnamefont
  {Gu}}, \bibinfo {author} {\bibfnamefont {S.}~\bibnamefont {Zhu}}, \bibinfo
  {author} {\bibfnamefont {X.}~\bibnamefont {Wang}}, \bibinfo {author}
  {\bibfnamefont {J.}~\bibnamefont {Hu}}, \ and\ \bibinfo {author}
  {\bibfnamefont {H.}~\bibnamefont {Chen}},\ }\href {\doibase
  10.1038/s42005-020-0347-x} {\bibfield  {journal} {\bibinfo  {journal}
  {Communications Physics}\ }\textbf {\bibinfo {volume} {3}},\ \bibinfo {pages}
  {84} (\bibinfo {year} {2020})}\BibitemShut {NoStop}%
\bibitem [{\citenamefont {Nomura}\ and\ \citenamefont
  {Arita}(2022)}]{Nomura2022}%
  \BibitemOpen
  \bibfield  {author} {\bibinfo {author} {\bibfnamefont {Y.}~\bibnamefont
  {Nomura}}\ and\ \bibinfo {author} {\bibfnamefont {R.}~\bibnamefont {Arita}},\
  }\href {\doibase 10.1088/1361-6633/ac5a60} {\bibfield  {journal} {\bibinfo
  {journal} {Reports on Progress in Physics}\ }\textbf {\bibinfo {volume}
  {85}},\ \bibinfo {pages} {052501} (\bibinfo {year} {2022})}\BibitemShut
  {NoStop}%
\bibitem [{\citenamefont {Wang}\ \emph {et~al.}(2022)\citenamefont {Wang},
  \citenamefont {Yang}, \citenamefont {Yang}, \citenamefont {Chen},
  \citenamefont {Zhang}, \citenamefont {Zhang}, \citenamefont {Zhu},
  \citenamefont {Uwatoko}, \citenamefont {Gu}, \citenamefont {Dong},
  \citenamefont {Sun}, \citenamefont {Jin},\ and\ \citenamefont
  {Cheng}}]{Wang2022}%
  \BibitemOpen
  \bibfield  {author} {\bibinfo {author} {\bibfnamefont {N.~N.}\ \bibnamefont
  {Wang}}, \bibinfo {author} {\bibfnamefont {M.~W.}\ \bibnamefont {Yang}},
  \bibinfo {author} {\bibfnamefont {Z.}~\bibnamefont {Yang}}, \bibinfo {author}
  {\bibfnamefont {K.~Y.}\ \bibnamefont {Chen}}, \bibinfo {author}
  {\bibfnamefont {H.}~\bibnamefont {Zhang}}, \bibinfo {author} {\bibfnamefont
  {Q.~H.}\ \bibnamefont {Zhang}}, \bibinfo {author} {\bibfnamefont {Z.~H.}\
  \bibnamefont {Zhu}}, \bibinfo {author} {\bibfnamefont {Y.}~\bibnamefont
  {Uwatoko}}, \bibinfo {author} {\bibfnamefont {L.}~\bibnamefont {Gu}},
  \bibinfo {author} {\bibfnamefont {X.~L.}\ \bibnamefont {Dong}}, \bibinfo
  {author} {\bibfnamefont {J.~P.}\ \bibnamefont {Sun}}, \bibinfo {author}
  {\bibfnamefont {K.~J.}\ \bibnamefont {Jin}}, \ and\ \bibinfo {author}
  {\bibfnamefont {J.~G.}\ \bibnamefont {Cheng}},\ }\href {\doibase
  10.1038/s41467-022-32065-x} {\bibfield  {journal} {\bibinfo  {journal}
  {Nature Communications}\ }\textbf {\bibinfo {volume} {13}},\ \bibinfo {pages}
  {4367} (\bibinfo {year} {2022})}\BibitemShut {NoStop}%
\bibitem [{\citenamefont {Gao}\ \emph {et~al.}(1994)\citenamefont {Gao},
  \citenamefont {Xue}, \citenamefont {Chen}, \citenamefont {Xiong},
  \citenamefont {Meng}, \citenamefont {Ramirez}, \citenamefont {Chu},
  \citenamefont {Eggert},\ and\ \citenamefont {Mao}}]{Gao1994}%
  \BibitemOpen
  \bibfield  {author} {\bibinfo {author} {\bibfnamefont {L.}~\bibnamefont
  {Gao}}, \bibinfo {author} {\bibfnamefont {Y.~Y.}\ \bibnamefont {Xue}},
  \bibinfo {author} {\bibfnamefont {F.}~\bibnamefont {Chen}}, \bibinfo {author}
  {\bibfnamefont {Q.}~\bibnamefont {Xiong}}, \bibinfo {author} {\bibfnamefont
  {R.~L.}\ \bibnamefont {Meng}}, \bibinfo {author} {\bibfnamefont
  {D.}~\bibnamefont {Ramirez}}, \bibinfo {author} {\bibfnamefont {C.~W.}\
  \bibnamefont {Chu}}, \bibinfo {author} {\bibfnamefont {J.~H.}\ \bibnamefont
  {Eggert}}, \ and\ \bibinfo {author} {\bibfnamefont {H.~K.}\ \bibnamefont
  {Mao}},\ }\href {\doibase 10.1103/PhysRevB.50.4260} {\bibfield  {journal}
  {\bibinfo  {journal} {Phys. Rev. B}\ }\textbf {\bibinfo {volume} {50}},\
  \bibinfo {pages} {4260} (\bibinfo {year} {1994})}\BibitemShut {NoStop}%
\bibitem [{\citenamefont {{Monteverde, M.}}\ \emph {et~al.}(2005)\citenamefont
  {{Monteverde, M.}}, \citenamefont {{Acha, C.}}, \citenamefont
  {{N\'u\~nez-Regueiro, M.}}, \citenamefont {{Pavlov, D. A.}}, \citenamefont
  {{Lokshin, K. A.}}, \citenamefont {{Putilin, S. N.}},\ and\ \citenamefont
  {{Antipov, E. V.}}}]{Monteverde2005}%
  \BibitemOpen
  \bibfield  {author} {\bibinfo {author} {\bibnamefont {{Monteverde, M.}}},
  \bibinfo {author} {\bibnamefont {{Acha, C.}}}, \bibinfo {author}
  {\bibnamefont {{N\'u\~nez-Regueiro, M.}}}, \bibinfo {author} {\bibnamefont
  {{Pavlov, D. A.}}}, \bibinfo {author} {\bibnamefont {{Lokshin, K. A.}}},
  \bibinfo {author} {\bibnamefont {{Putilin, S. N.}}}, \ and\ \bibinfo {author}
  {\bibnamefont {{Antipov, E. V.}}},\ }\href {\doibase
  10.1209/epl/i2005-10247-3} {\bibfield  {journal} {\bibinfo  {journal}
  {Europhys. Lett.}\ }\textbf {\bibinfo {volume} {72}},\ \bibinfo {pages} {458}
  (\bibinfo {year} {2005})}\BibitemShut {NoStop}%
\bibitem [{\citenamefont {Been}\ \emph {et~al.}(2021)\citenamefont {Been},
  \citenamefont {Lee}, \citenamefont {Hwang}, \citenamefont {Cui},
  \citenamefont {Zaanen}, \citenamefont {Devereaux}, \citenamefont {Moritz},\
  and\ \citenamefont {Jia}}]{Been2021}%
  \BibitemOpen
  \bibfield  {author} {\bibinfo {author} {\bibfnamefont {E.}~\bibnamefont
  {Been}}, \bibinfo {author} {\bibfnamefont {W.-S.}\ \bibnamefont {Lee}},
  \bibinfo {author} {\bibfnamefont {H.~Y.}\ \bibnamefont {Hwang}}, \bibinfo
  {author} {\bibfnamefont {Y.}~\bibnamefont {Cui}}, \bibinfo {author}
  {\bibfnamefont {J.}~\bibnamefont {Zaanen}}, \bibinfo {author} {\bibfnamefont
  {T.}~\bibnamefont {Devereaux}}, \bibinfo {author} {\bibfnamefont
  {B.}~\bibnamefont {Moritz}}, \ and\ \bibinfo {author} {\bibfnamefont
  {C.}~\bibnamefont {Jia}},\ }\href {\doibase 10.1103/PhysRevX.11.011050}
  {\bibfield  {journal} {\bibinfo  {journal} {Phys. Rev. X}\ }\textbf {\bibinfo
  {volume} {11}},\ \bibinfo {pages} {011050} (\bibinfo {year}
  {2021})}\BibitemShut {NoStop}%
\bibitem [{\citenamefont {Bernardini}\ \emph {et~al.}(2022)\citenamefont
  {Bernardini}, \citenamefont {Bosin},\ and\ \citenamefont
  {Cano}}]{Bernardini2022}%
  \BibitemOpen
  \bibfield  {author} {\bibinfo {author} {\bibfnamefont {F.}~\bibnamefont
  {Bernardini}}, \bibinfo {author} {\bibfnamefont {A.}~\bibnamefont {Bosin}}, \
  and\ \bibinfo {author} {\bibfnamefont {A.}~\bibnamefont {Cano}},\ }\href
  {\doibase 10.1103/PhysRevMaterials.6.044807} {\bibfield  {journal} {\bibinfo
  {journal} {Phys. Rev. Materials}\ }\textbf {\bibinfo {volume} {6}},\ \bibinfo
  {pages} {044807} (\bibinfo {year} {2022})}\BibitemShut {NoStop}%
\bibitem [{\citenamefont {Si}\ \emph {et~al.}(2020)\citenamefont {Si},
  \citenamefont {Xiao}, \citenamefont {Kaufmann}, \citenamefont {Tomczak},
  \citenamefont {Lu}, \citenamefont {Zhong},\ and\ \citenamefont
  {Held}}]{Si2020}%
  \BibitemOpen
  \bibfield  {author} {\bibinfo {author} {\bibfnamefont {L.}~\bibnamefont
  {Si}}, \bibinfo {author} {\bibfnamefont {W.}~\bibnamefont {Xiao}}, \bibinfo
  {author} {\bibfnamefont {J.}~\bibnamefont {Kaufmann}}, \bibinfo {author}
  {\bibfnamefont {J.~M.}\ \bibnamefont {Tomczak}}, \bibinfo {author}
  {\bibfnamefont {Y.}~\bibnamefont {Lu}}, \bibinfo {author} {\bibfnamefont
  {Z.}~\bibnamefont {Zhong}}, \ and\ \bibinfo {author} {\bibfnamefont
  {K.}~\bibnamefont {Held}},\ }\href {\doibase 10.1103/PhysRevLett.124.166402}
  {\bibfield  {journal} {\bibinfo  {journal} {Phys. Rev. Lett.}\ }\textbf
  {\bibinfo {volume} {124}},\ \bibinfo {pages} {166402} (\bibinfo {year}
  {2020})}\BibitemShut {NoStop}%
\bibitem [{\citenamefont {Malyi}\ \emph {et~al.}(2022)\citenamefont {Malyi},
  \citenamefont {Varignon},\ and\ \citenamefont {Zunger}}]{Malyi2022}%
  \BibitemOpen
  \bibfield  {author} {\bibinfo {author} {\bibfnamefont {O.~I.}\ \bibnamefont
  {Malyi}}, \bibinfo {author} {\bibfnamefont {J.}~\bibnamefont {Varignon}}, \
  and\ \bibinfo {author} {\bibfnamefont {A.}~\bibnamefont {Zunger}},\ }\href
  {\doibase 10.1103/PhysRevB.105.014106} {\bibfield  {journal} {\bibinfo
  {journal} {Phys. Rev. B}\ }\textbf {\bibinfo {volume} {105}},\ \bibinfo
  {pages} {014106} (\bibinfo {year} {2022})}\BibitemShut {NoStop}%
\bibitem [{\citenamefont {Anisimov}\ \emph {et~al.}(1997)\citenamefont
  {Anisimov}, \citenamefont {Aryasetiawan},\ and\ \citenamefont
  {Lichtenstein}}]{Anisimov1997}%
  \BibitemOpen
  \bibfield  {author} {\bibinfo {author} {\bibfnamefont {V.~I.}\ \bibnamefont
  {Anisimov}}, \bibinfo {author} {\bibfnamefont {F.}~\bibnamefont
  {Aryasetiawan}}, \ and\ \bibinfo {author} {\bibfnamefont {A.~I.}\
  \bibnamefont {Lichtenstein}},\ }\href {\doibase 10.1088/0953-8984/9/4/002}
  {\bibfield  {journal} {\bibinfo  {journal} {Journal of Physics: Condensed
  Matter}\ }\textbf {\bibinfo {volume} {9}},\ \bibinfo {pages} {767} (\bibinfo
  {year} {1997})}\BibitemShut {NoStop}%
\bibitem [{\citenamefont {Rohringer}\ \emph {et~al.}(2018)\citenamefont
  {Rohringer}, \citenamefont {Hafermann}, \citenamefont {Toschi}, \citenamefont
  {Katanin}, \citenamefont {Antipov}, \citenamefont {Katsnelson}, \citenamefont
  {Lichtenstein}, \citenamefont {Rubtsov},\ and\ \citenamefont
  {Held}}]{Rohringer2018}%
  \BibitemOpen
  \bibfield  {author} {\bibinfo {author} {\bibfnamefont {G.}~\bibnamefont
  {Rohringer}}, \bibinfo {author} {\bibfnamefont {H.}~\bibnamefont
  {Hafermann}}, \bibinfo {author} {\bibfnamefont {A.}~\bibnamefont {Toschi}},
  \bibinfo {author} {\bibfnamefont {A.~A.}\ \bibnamefont {Katanin}}, \bibinfo
  {author} {\bibfnamefont {A.~E.}\ \bibnamefont {Antipov}}, \bibinfo {author}
  {\bibfnamefont {M.~I.}\ \bibnamefont {Katsnelson}}, \bibinfo {author}
  {\bibfnamefont {A.~I.}\ \bibnamefont {Lichtenstein}}, \bibinfo {author}
  {\bibfnamefont {A.~N.}\ \bibnamefont {Rubtsov}}, \ and\ \bibinfo {author}
  {\bibfnamefont {K.}~\bibnamefont {Held}},\ }\href {\doibase
  10.1103/RevModPhys.90.025003} {\bibfield  {journal} {\bibinfo  {journal}
  {Rev. Mod. Phys.}\ }\textbf {\bibinfo {volume} {90}},\ \bibinfo {pages}
  {025003} (\bibinfo {year} {2018})}\BibitemShut {NoStop}%
\bibitem [{\citenamefont {Li}\ \emph {et~al.}(2020)\citenamefont {Li},
  \citenamefont {Wang}, \citenamefont {Lee}, \citenamefont {Harvey},
  \citenamefont {Osada}, \citenamefont {Goodge}, \citenamefont {Kourkoutis},\
  and\ \citenamefont {Hwang}}]{Li2020}%
  \BibitemOpen
  \bibfield  {author} {\bibinfo {author} {\bibfnamefont {D.}~\bibnamefont
  {Li}}, \bibinfo {author} {\bibfnamefont {B.~Y.}\ \bibnamefont {Wang}},
  \bibinfo {author} {\bibfnamefont {K.}~\bibnamefont {Lee}}, \bibinfo {author}
  {\bibfnamefont {S.~P.}\ \bibnamefont {Harvey}}, \bibinfo {author}
  {\bibfnamefont {M.}~\bibnamefont {Osada}}, \bibinfo {author} {\bibfnamefont
  {B.~H.}\ \bibnamefont {Goodge}}, \bibinfo {author} {\bibfnamefont {L.~F.}\
  \bibnamefont {Kourkoutis}}, \ and\ \bibinfo {author} {\bibfnamefont {H.~Y.}\
  \bibnamefont {Hwang}},\ }\href {\doibase 10.1103/PhysRevLett.125.027001}
  {\bibfield  {journal} {\bibinfo  {journal} {Phys. Rev. Lett.}\ }\textbf
  {\bibinfo {volume} {125}},\ \bibinfo {pages} {027001} (\bibinfo {year}
  {2020})}\BibitemShut {NoStop}%
\bibitem [{\citenamefont {Zeng}\ \emph {et~al.}(2020)\citenamefont {Zeng},
  \citenamefont {Tang}, \citenamefont {Yin}, \citenamefont {Li}, \citenamefont
  {Li}, \citenamefont {Huang}, \citenamefont {Hu}, \citenamefont {Liu},
  \citenamefont {Omar}, \citenamefont {Jani}, \citenamefont {Lim},
  \citenamefont {Han}, \citenamefont {Wan}, \citenamefont {Yang}, \citenamefont
  {Pennycook}, \citenamefont {Wee},\ and\ \citenamefont {Ariando}}]{Zeng2020}%
  \BibitemOpen
  \bibfield  {author} {\bibinfo {author} {\bibfnamefont {S.}~\bibnamefont
  {Zeng}}, \bibinfo {author} {\bibfnamefont {C.~S.}\ \bibnamefont {Tang}},
  \bibinfo {author} {\bibfnamefont {X.}~\bibnamefont {Yin}}, \bibinfo {author}
  {\bibfnamefont {C.}~\bibnamefont {Li}}, \bibinfo {author} {\bibfnamefont
  {M.}~\bibnamefont {Li}}, \bibinfo {author} {\bibfnamefont {Z.}~\bibnamefont
  {Huang}}, \bibinfo {author} {\bibfnamefont {J.}~\bibnamefont {Hu}}, \bibinfo
  {author} {\bibfnamefont {W.}~\bibnamefont {Liu}}, \bibinfo {author}
  {\bibfnamefont {G.~J.}\ \bibnamefont {Omar}}, \bibinfo {author}
  {\bibfnamefont {H.}~\bibnamefont {Jani}}, \bibinfo {author} {\bibfnamefont
  {Z.~S.}\ \bibnamefont {Lim}}, \bibinfo {author} {\bibfnamefont
  {K.}~\bibnamefont {Han}}, \bibinfo {author} {\bibfnamefont {D.}~\bibnamefont
  {Wan}}, \bibinfo {author} {\bibfnamefont {P.}~\bibnamefont {Yang}}, \bibinfo
  {author} {\bibfnamefont {S.~J.}\ \bibnamefont {Pennycook}}, \bibinfo {author}
  {\bibfnamefont {A.~T.~S.}\ \bibnamefont {Wee}}, \ and\ \bibinfo {author}
  {\bibfnamefont {A.}~\bibnamefont {Ariando}},\ }\href {\doibase
  10.1103/PhysRevLett.125.147003} {\bibfield  {journal} {\bibinfo  {journal}
  {Phys. Rev. Lett.}\ }\textbf {\bibinfo {volume} {125}},\ \bibinfo {pages}
  {147003} (\bibinfo {year} {2020})}\BibitemShut {NoStop}%
\bibitem [{\citenamefont {Biermann}\ \emph {et~al.}(2003)\citenamefont
  {Biermann}, \citenamefont {Aryasetiawan},\ and\ \citenamefont
  {Georges}}]{Biermann2003}%
  \BibitemOpen
  \bibfield  {author} {\bibinfo {author} {\bibfnamefont {S.}~\bibnamefont
  {Biermann}}, \bibinfo {author} {\bibfnamefont {F.}~\bibnamefont
  {Aryasetiawan}}, \ and\ \bibinfo {author} {\bibfnamefont {A.}~\bibnamefont
  {Georges}},\ }\href {\doibase 10.1103/PhysRevLett.90.086402} {\bibfield
  {journal} {\bibinfo  {journal} {{Physical Review Letters}}\ }\textbf
  {\bibinfo {volume} {90}},\ \bibinfo {pages} {086402} (\bibinfo {year}
  {2003})}\BibitemShut {NoStop}%
\bibitem [{\citenamefont {Ayral}\ \emph {et~al.}(2013)\citenamefont {Ayral},
  \citenamefont {Biermann},\ and\ \citenamefont {Werner}}]{Ayral2013}%
  \BibitemOpen
  \bibfield  {author} {\bibinfo {author} {\bibfnamefont {T.}~\bibnamefont
  {Ayral}}, \bibinfo {author} {\bibfnamefont {S.}~\bibnamefont {Biermann}}, \
  and\ \bibinfo {author} {\bibfnamefont {P.}~\bibnamefont {Werner}},\ }\href
  {\doibase 10.1103/PhysRevB.87.125149} {\bibfield  {journal} {\bibinfo
  {journal} {Phys. Rev. B}\ }\textbf {\bibinfo {volume} {87}},\ \bibinfo
  {pages} {125149} (\bibinfo {year} {2013})}\BibitemShut {NoStop}%
\bibitem [{\citenamefont {Boehnke}\ \emph {et~al.}(2016)\citenamefont
  {Boehnke}, \citenamefont {Nilsson}, \citenamefont {Aryasetiawan},\ and\
  \citenamefont {Werner}}]{Boehnke2016}%
  \BibitemOpen
  \bibfield  {author} {\bibinfo {author} {\bibfnamefont {L.}~\bibnamefont
  {Boehnke}}, \bibinfo {author} {\bibfnamefont {F.}~\bibnamefont {Nilsson}},
  \bibinfo {author} {\bibfnamefont {F.}~\bibnamefont {Aryasetiawan}}, \ and\
  \bibinfo {author} {\bibfnamefont {P.}~\bibnamefont {Werner}},\ }\href
  {\doibase 10.1103/PhysRevB.94.201106} {\bibfield  {journal} {\bibinfo
  {journal} {Phys. Rev. B}\ }\textbf {\bibinfo {volume} {94}},\ \bibinfo
  {pages} {201106} (\bibinfo {year} {2016})}\BibitemShut {NoStop}%
\bibitem [{\citenamefont {Nilsson}\ \emph {et~al.}(2017)\citenamefont
  {Nilsson}, \citenamefont {Boehnke}, \citenamefont {Werner},\ and\
  \citenamefont {Aryasetiawan}}]{Nilsson2017}%
  \BibitemOpen
  \bibfield  {author} {\bibinfo {author} {\bibfnamefont {F.}~\bibnamefont
  {Nilsson}}, \bibinfo {author} {\bibfnamefont {L.}~\bibnamefont {Boehnke}},
  \bibinfo {author} {\bibfnamefont {P.}~\bibnamefont {Werner}}, \ and\ \bibinfo
  {author} {\bibfnamefont {F.}~\bibnamefont {Aryasetiawan}},\ }\href {\doibase
  10.1103/PhysRevMaterials.1.043803} {\bibfield  {journal} {\bibinfo  {journal}
  {Phys. Rev. Materials}\ }\textbf {\bibinfo {volume} {1}},\ \bibinfo {pages}
  {043803} (\bibinfo {year} {2017})}\BibitemShut {NoStop}%
\bibitem [{\citenamefont {Hohenberg}\ and\ \citenamefont
  {Kohn}(1964)}]{Hohenberg1964}%
  \BibitemOpen
  \bibfield  {author} {\bibinfo {author} {\bibfnamefont {P.}~\bibnamefont
  {Hohenberg}}\ and\ \bibinfo {author} {\bibfnamefont {W.}~\bibnamefont
  {Kohn}},\ }\href {\doibase 10.1103/PhysRev.136.B864} {\bibfield  {journal}
  {\bibinfo  {journal} {Phys. Rev.}\ }\textbf {\bibinfo {volume} {136}},\
  \bibinfo {pages} {B864} (\bibinfo {year} {1964})}\BibitemShut {NoStop}%
\bibitem [{\citenamefont {Kohn}\ and\ \citenamefont {Sham}(1965)}]{Kohn1965}%
  \BibitemOpen
  \bibfield  {author} {\bibinfo {author} {\bibfnamefont {W.}~\bibnamefont
  {Kohn}}\ and\ \bibinfo {author} {\bibfnamefont {L.~J.}\ \bibnamefont
  {Sham}},\ }\href {\doibase 10.1103/PhysRev.140.A1133} {\bibfield  {journal}
  {\bibinfo  {journal} {Phys. Rev.}\ }\textbf {\bibinfo {volume} {140}},\
  \bibinfo {pages} {A1133} (\bibinfo {year} {1965})}\BibitemShut {NoStop}%
\bibitem [{\citenamefont {Perdew}\ \emph {et~al.}(1996)\citenamefont {Perdew},
  \citenamefont {Burke},\ and\ \citenamefont {Ernzerhof}}]{Perdew1996}%
  \BibitemOpen
  \bibfield  {author} {\bibinfo {author} {\bibfnamefont {J.~P.}\ \bibnamefont
  {Perdew}}, \bibinfo {author} {\bibfnamefont {K.}~\bibnamefont {Burke}}, \
  and\ \bibinfo {author} {\bibfnamefont {M.}~\bibnamefont {Ernzerhof}},\ }\href
  {\doibase 10.1103/PhysRevLett.77.3865} {\bibfield  {journal} {\bibinfo
  {journal} {Phys. Rev. Lett.}\ }\textbf {\bibinfo {volume} {77}},\ \bibinfo
  {pages} {3865} (\bibinfo {year} {1996})}\BibitemShut {NoStop}%
\bibitem [{\citenamefont {{The FLEUR group}}()}]{Fleurcode}%
  \BibitemOpen
  \bibfield  {author} {\bibinfo {author} {\bibnamefont {{The FLEUR group}}},\
  }\href@noop {} {\enquote {\bibinfo {title} {{The FLEUR project}},}\ }\bibinfo
  {howpublished} {\url{http://www.flapw.de}}\BibitemShut {NoStop}%
\bibitem [{\citenamefont {Bellaiche}\ and\ \citenamefont
  {Vanderbilt}(2000)}]{Bellaiche2000}%
  \BibitemOpen
  \bibfield  {author} {\bibinfo {author} {\bibfnamefont {L.}~\bibnamefont
  {Bellaiche}}\ and\ \bibinfo {author} {\bibfnamefont {D.}~\bibnamefont
  {Vanderbilt}},\ }\href {\doibase 10.1103/PhysRevB.61.7877} {\bibfield
  {journal} {\bibinfo  {journal} {Phys. Rev. B}\ }\textbf {\bibinfo {volume}
  {61}},\ \bibinfo {pages} {7877} (\bibinfo {year} {2000})}\BibitemShut
  {NoStop}%
\bibitem [{Note1()}]{Note1}%
  \BibitemOpen
  \bibinfo {note} {In this work, they also varied the in-plane lattice constant
  on the order of 1\% to follow more closely the effect of the change in the
  rare-earth element.}\BibitemShut {Stop}%
\bibitem [{\citenamefont {Vinet}\ \emph {et~al.}(1986)\citenamefont {Vinet},
  \citenamefont {Ferrante}, \citenamefont {Smith},\ and\ \citenamefont
  {Rose}}]{Vinet1986}%
  \BibitemOpen
  \bibfield  {author} {\bibinfo {author} {\bibfnamefont {P.}~\bibnamefont
  {Vinet}}, \bibinfo {author} {\bibfnamefont {J.}~\bibnamefont {Ferrante}},
  \bibinfo {author} {\bibfnamefont {J.~R.}\ \bibnamefont {Smith}}, \ and\
  \bibinfo {author} {\bibfnamefont {J.~H.}\ \bibnamefont {Rose}},\ }\href
  {\doibase 10.1088/0022-3719/19/20/001} {\bibfield  {journal} {\bibinfo
  {journal} {Journal of Physics C: Solid State Physics}\ }\textbf {\bibinfo
  {volume} {19}},\ \bibinfo {pages} {L467} (\bibinfo {year}
  {1986})}\BibitemShut {NoStop}%
\bibitem [{\citenamefont {Aryasetiawan}\ \emph {et~al.}(2004)\citenamefont
  {Aryasetiawan}, \citenamefont {Imada}, \citenamefont {Georges}, \citenamefont
  {Kotliar}, \citenamefont {Biermann},\ and\ \citenamefont
  {Lichtenstein}}]{Aryasetiawan2004}%
  \BibitemOpen
  \bibfield  {author} {\bibinfo {author} {\bibfnamefont {F.}~\bibnamefont
  {Aryasetiawan}}, \bibinfo {author} {\bibfnamefont {M.}~\bibnamefont {Imada}},
  \bibinfo {author} {\bibfnamefont {A.}~\bibnamefont {Georges}}, \bibinfo
  {author} {\bibfnamefont {G.}~\bibnamefont {Kotliar}}, \bibinfo {author}
  {\bibfnamefont {S.}~\bibnamefont {Biermann}}, \ and\ \bibinfo {author}
  {\bibfnamefont {A.~I.}\ \bibnamefont {Lichtenstein}},\ }\href {\doibase
  10.1103/PhysRevB.70.195104} {\bibfield  {journal} {\bibinfo  {journal} {Phys.
  Rev. B}\ }\textbf {\bibinfo {volume} {70}},\ \bibinfo {pages} {195104}
  (\bibinfo {year} {2004})}\BibitemShut {NoStop}%
\bibitem [{\citenamefont {Petocchi}\ \emph
  {et~al.}(2020{\natexlab{b}})\citenamefont {Petocchi}, \citenamefont
  {Nilsson}, \citenamefont {Aryasetiawan},\ and\ \citenamefont
  {Werner}}]{Petocchi2020a}%
  \BibitemOpen
  \bibfield  {author} {\bibinfo {author} {\bibfnamefont {F.}~\bibnamefont
  {Petocchi}}, \bibinfo {author} {\bibfnamefont {F.}~\bibnamefont {Nilsson}},
  \bibinfo {author} {\bibfnamefont {F.}~\bibnamefont {Aryasetiawan}}, \ and\
  \bibinfo {author} {\bibfnamefont {P.}~\bibnamefont {Werner}},\ }\href
  {\doibase 10.1103/PhysRevResearch.2.013191} {\bibfield  {journal} {\bibinfo
  {journal} {Phys. Rev. Research}\ }\textbf {\bibinfo {volume} {2}},\ \bibinfo
  {pages} {013191} (\bibinfo {year} {2020}{\natexlab{b}})}\BibitemShut
  {NoStop}%
\bibitem [{\citenamefont {Petocchi}\ \emph {et~al.}(2021)\citenamefont
  {Petocchi}, \citenamefont {Christiansson},\ and\ \citenamefont
  {Werner}}]{Petocchi2021}%
  \BibitemOpen
  \bibfield  {author} {\bibinfo {author} {\bibfnamefont {F.}~\bibnamefont
  {Petocchi}}, \bibinfo {author} {\bibfnamefont {V.}~\bibnamefont
  {Christiansson}}, \ and\ \bibinfo {author} {\bibfnamefont {P.}~\bibnamefont
  {Werner}},\ }\href {\doibase 10.1103/PhysRevB.104.195146} {\bibfield
  {journal} {\bibinfo  {journal} {Phys. Rev. B}\ }\textbf {\bibinfo {volume}
  {104}},\ \bibinfo {pages} {195146} (\bibinfo {year} {2021})}\BibitemShut
  {NoStop}%
\bibitem [{\citenamefont {Christiansson}\ \emph {et~al.}(2022)\citenamefont
  {Christiansson}, \citenamefont {Petocchi},\ and\ \citenamefont
  {Werner}}]{Christiansson2022}%
  \BibitemOpen
  \bibfield  {author} {\bibinfo {author} {\bibfnamefont {V.}~\bibnamefont
  {Christiansson}}, \bibinfo {author} {\bibfnamefont {F.}~\bibnamefont
  {Petocchi}}, \ and\ \bibinfo {author} {\bibfnamefont {P.}~\bibnamefont
  {Werner}},\ }\href {\doibase 10.1103/PhysRevB.105.174513} {\bibfield
  {journal} {\bibinfo  {journal} {Phys. Rev. B}\ }\textbf {\bibinfo {volume}
  {105}},\ \bibinfo {pages} {174513} (\bibinfo {year} {2022})}\BibitemShut
  {NoStop}%
\bibitem [{\citenamefont {Marzari}\ and\ \citenamefont
  {Vanderbilt}(1997)}]{Marzari1997}%
  \BibitemOpen
  \bibfield  {author} {\bibinfo {author} {\bibfnamefont {N.}~\bibnamefont
  {Marzari}}\ and\ \bibinfo {author} {\bibfnamefont {D.}~\bibnamefont
  {Vanderbilt}},\ }\href {\doibase 10.1103/PhysRevB.56.12847} {\bibfield
  {journal} {\bibinfo  {journal} {{Phys. Rev. B}}\ }\textbf {\bibinfo {volume}
  {56}},\ \bibinfo {pages} {12847} (\bibinfo {year} {1997})}\BibitemShut
  {NoStop}%
\bibitem [{\citenamefont {Mostofi}\ \emph {et~al.}(2008)\citenamefont
  {Mostofi}, \citenamefont {Yates}, \citenamefont {Lee}, \citenamefont {Souza},
  \citenamefont {Vanderbilt},\ and\ \citenamefont {Marzari}}]{Mostofi2008}%
  \BibitemOpen
  \bibfield  {author} {\bibinfo {author} {\bibfnamefont {A.~A.}\ \bibnamefont
  {Mostofi}}, \bibinfo {author} {\bibfnamefont {J.~R.}\ \bibnamefont {Yates}},
  \bibinfo {author} {\bibfnamefont {Y.-S.}\ \bibnamefont {Lee}}, \bibinfo
  {author} {\bibfnamefont {I.}~\bibnamefont {Souza}}, \bibinfo {author}
  {\bibfnamefont {D.}~\bibnamefont {Vanderbilt}}, \ and\ \bibinfo {author}
  {\bibfnamefont {N.}~\bibnamefont {Marzari}},\ }\href {\doibase
  {http://dx.doi.org/10.1016/j.cpc.2007.11.016}} {\bibfield  {journal}
  {\bibinfo  {journal} {{Computer Physics Communications }}\ }\textbf {\bibinfo
  {volume} {178}},\ \bibinfo {pages} {685} (\bibinfo {year}
  {2008})}\BibitemShut {NoStop}%
\bibitem [{\citenamefont {Hedin}(1965)}]{Hedin1965}%
  \BibitemOpen
  \bibfield  {author} {\bibinfo {author} {\bibfnamefont {L.}~\bibnamefont
  {Hedin}},\ }\href {\doibase 10.1103/PhysRev.139.A796} {\bibfield  {journal}
  {\bibinfo  {journal} {Phys. Rev.}\ }\textbf {\bibinfo {volume} {139}},\
  \bibinfo {pages} {A796} (\bibinfo {year} {1965})}\BibitemShut {NoStop}%
\bibitem [{\citenamefont {Friedrich}\ \emph {et~al.}(2010)\citenamefont
  {Friedrich}, \citenamefont {Bl{\"{u}}gel},\ and\ \citenamefont
  {Schindlmayr}}]{Friedrich2010}%
  \BibitemOpen
  \bibfield  {author} {\bibinfo {author} {\bibfnamefont {C.}~\bibnamefont
  {Friedrich}}, \bibinfo {author} {\bibfnamefont {S.}~\bibnamefont
  {Bl{\"{u}}gel}}, \ and\ \bibinfo {author} {\bibfnamefont {A.}~\bibnamefont
  {Schindlmayr}},\ }\href {\doibase 10.1103/PhysRevB.81.125102} {\bibfield
  {journal} {\bibinfo  {journal} {{Phys. Rev. B}}\ }\textbf {\bibinfo {volume}
  {81}},\ \bibinfo {pages} {125102} (\bibinfo {year} {2010})}\BibitemShut
  {NoStop}%
\bibitem [{\citenamefont {Georges}\ \emph {et~al.}(1996)\citenamefont
  {Georges}, \citenamefont {Kotliar}, \citenamefont {Krauth},\ and\
  \citenamefont {Rozenberg}}]{Georges1996}%
  \BibitemOpen
  \bibfield  {author} {\bibinfo {author} {\bibfnamefont {A.}~\bibnamefont
  {Georges}}, \bibinfo {author} {\bibfnamefont {G.}~\bibnamefont {Kotliar}},
  \bibinfo {author} {\bibfnamefont {W.}~\bibnamefont {Krauth}}, \ and\ \bibinfo
  {author} {\bibfnamefont {M.~J.}\ \bibnamefont {Rozenberg}},\ }\href {\doibase
  10.1103/RevModPhys.68.13} {\bibfield  {journal} {\bibinfo  {journal} {Rev.
  Mod. Phys.}\ }\textbf {\bibinfo {volume} {68}},\ \bibinfo {pages} {13}
  (\bibinfo {year} {1996})}\BibitemShut {NoStop}%
\bibitem [{\citenamefont {Sun}\ and\ \citenamefont {Kotliar}(2002)}]{Sun2002}%
  \BibitemOpen
  \bibfield  {author} {\bibinfo {author} {\bibfnamefont {P.}~\bibnamefont
  {Sun}}\ and\ \bibinfo {author} {\bibfnamefont {G.}~\bibnamefont {Kotliar}},\
  }\href {\doibase 10.1103/PhysRevB.66.085120} {\bibfield  {journal} {\bibinfo
  {journal} {Phys. Rev. B}\ }\textbf {\bibinfo {volume} {66}},\ \bibinfo
  {pages} {085120} (\bibinfo {year} {2002})}\BibitemShut {NoStop}%
\bibitem [{\citenamefont {Werner}\ \emph {et~al.}(2006)\citenamefont {Werner},
  \citenamefont {Comanac}, \citenamefont {de' Medici}, \citenamefont {Troyer},\
  and\ \citenamefont {Millis}}]{Werner2006}%
  \BibitemOpen
  \bibfield  {author} {\bibinfo {author} {\bibfnamefont {P.}~\bibnamefont
  {Werner}}, \bibinfo {author} {\bibfnamefont {A.}~\bibnamefont {Comanac}},
  \bibinfo {author} {\bibfnamefont {L.}~\bibnamefont {de' Medici}}, \bibinfo
  {author} {\bibfnamefont {M.}~\bibnamefont {Troyer}}, \ and\ \bibinfo {author}
  {\bibfnamefont {A.~J.}\ \bibnamefont {Millis}},\ }\href {\doibase
  10.1103/PhysRevLett.97.076405} {\bibfield  {journal} {\bibinfo  {journal}
  {Phys. Rev. Lett.}\ }\textbf {\bibinfo {volume} {97}},\ \bibinfo {pages}
  {076405} (\bibinfo {year} {2006})}\BibitemShut {NoStop}%
\bibitem [{\citenamefont {Hafermann}\ \emph {et~al.}(2013)\citenamefont
  {Hafermann}, \citenamefont {Werner},\ and\ \citenamefont
  {Gull}}]{Hafermann2013}%
  \BibitemOpen
  \bibfield  {author} {\bibinfo {author} {\bibfnamefont {H.}~\bibnamefont
  {Hafermann}}, \bibinfo {author} {\bibfnamefont {P.}~\bibnamefont {Werner}}, \
  and\ \bibinfo {author} {\bibfnamefont {E.}~\bibnamefont {Gull}},\ }\href
  {\doibase http://dx.doi.org/10.1016/j.cpc.2012.12.013} {\bibfield  {journal}
  {\bibinfo  {journal} {Comput. Phys. Commun.}\ }\textbf {\bibinfo {volume}
  {184}},\ \bibinfo {pages} {1280 } (\bibinfo {year} {2013})}\BibitemShut
  {NoStop}%
\bibitem [{\citenamefont {Werner}\ and\ \citenamefont
  {Millis}(2010)}]{Werner2010}%
  \BibitemOpen
  \bibfield  {author} {\bibinfo {author} {\bibfnamefont {P.}~\bibnamefont
  {Werner}}\ and\ \bibinfo {author} {\bibfnamefont {A.~J.}\ \bibnamefont
  {Millis}},\ }\href {\doibase 10.1103/PhysRevLett.104.146401} {\bibfield
  {journal} {\bibinfo  {journal} {Phys. Rev. Lett.}\ }\textbf {\bibinfo
  {volume} {104}},\ \bibinfo {pages} {146401} (\bibinfo {year}
  {2010})}\BibitemShut {NoStop}%
\bibitem [{\citenamefont {Tomczak}\ \emph {et~al.}(2009)\citenamefont
  {Tomczak}, \citenamefont {Miyake}, \citenamefont {Sakuma},\ and\
  \citenamefont {Aryasetiawan}}]{Tomczak2009}%
  \BibitemOpen
  \bibfield  {author} {\bibinfo {author} {\bibfnamefont {J.~M.}\ \bibnamefont
  {Tomczak}}, \bibinfo {author} {\bibfnamefont {T.}~\bibnamefont {Miyake}},
  \bibinfo {author} {\bibfnamefont {R.}~\bibnamefont {Sakuma}}, \ and\ \bibinfo
  {author} {\bibfnamefont {F.}~\bibnamefont {Aryasetiawan}},\ }\href {\doibase
  10.1103/PhysRevB.79.235133} {\bibfield  {journal} {\bibinfo  {journal} {Phys.
  Rev. B}\ }\textbf {\bibinfo {volume} {79}},\ \bibinfo {pages} {235133}
  (\bibinfo {year} {2009})}\BibitemShut {NoStop}%
\bibitem [{\citenamefont {Held}\ \emph {et~al.}(2022)\citenamefont {Held},
  \citenamefont {Si}, \citenamefont {Worm}, \citenamefont {Janson},
  \citenamefont {Arita}, \citenamefont {Zhong}, \citenamefont {Tomczak},\ and\
  \citenamefont {Kitatani}}]{Held2022}%
  \BibitemOpen
  \bibfield  {author} {\bibinfo {author} {\bibfnamefont {K.}~\bibnamefont
  {Held}}, \bibinfo {author} {\bibfnamefont {L.}~\bibnamefont {Si}}, \bibinfo
  {author} {\bibfnamefont {P.}~\bibnamefont {Worm}}, \bibinfo {author}
  {\bibfnamefont {O.}~\bibnamefont {Janson}}, \bibinfo {author} {\bibfnamefont
  {R.}~\bibnamefont {Arita}}, \bibinfo {author} {\bibfnamefont
  {Z.}~\bibnamefont {Zhong}}, \bibinfo {author} {\bibfnamefont {J.~M.}\
  \bibnamefont {Tomczak}}, \ and\ \bibinfo {author} {\bibfnamefont
  {M.}~\bibnamefont {Kitatani}},\ }\href {\doibase 10.3389/fphy.2021.810394}
  {\bibfield  {journal} {\bibinfo  {journal} {Frontiers in Physics}\ }\textbf
  {\bibinfo {volume} {9}} (\bibinfo {year} {2022}),\
  10.3389/fphy.2021.810394}\BibitemShut {NoStop}%
\bibitem [{\citenamefont {Sakakibara}\ \emph {et~al.}(2010)\citenamefont
  {Sakakibara}, \citenamefont {Usui}, \citenamefont {Kuroki}, \citenamefont
  {Arita},\ and\ \citenamefont {Aoki}}]{Sakakibara2010}%
  \BibitemOpen
  \bibfield  {author} {\bibinfo {author} {\bibfnamefont {H.}~\bibnamefont
  {Sakakibara}}, \bibinfo {author} {\bibfnamefont {H.}~\bibnamefont {Usui}},
  \bibinfo {author} {\bibfnamefont {K.}~\bibnamefont {Kuroki}}, \bibinfo
  {author} {\bibfnamefont {R.}~\bibnamefont {Arita}}, \ and\ \bibinfo {author}
  {\bibfnamefont {H.}~\bibnamefont {Aoki}},\ }\href {\doibase
  10.1103/PhysRevLett.105.057003} {\bibfield  {journal} {\bibinfo  {journal}
  {Phys. Rev. Lett.}\ }\textbf {\bibinfo {volume} {105}},\ \bibinfo {pages}
  {057003} (\bibinfo {year} {2010})}\BibitemShut {NoStop}%
\end{thebibliography}%

\end{document}